\documentclass[11pt]{article}

\usepackage{mathpazo}
\usepackage{times}

\usepackage{amsthm}
\usepackage{amsmath}  
\usepackage{amsfonts}
\usepackage{amssymb}  
\usepackage{mathrsfs} 
\usepackage{mathtools}
\usepackage{cite}     

\usepackage{tikz}
\usepackage{fullpage}

\usepackage{xfrac}
\usepackage{epstopdf}
\usepackage{hhline}
\usepackage{etoolbox}
\usepackage[singlelinecheck=false]{caption} 
\usepackage{bm}
\usepackage{url}
\usepackage{hyperref}
\hypersetup{colorlinks=false}

\title{\Huge $\,$\\[-7.20ex]
Improved Schemes for Asymptotically\\
Optimal Repair of MDS Codes\\[0.90ex]}

\author{
Ameera Chowdhury\\
   \small University of California San Diego\vspace*{-0.72ex}\\
   \small 9500 Gilman Drive, La Jolla, CA\,92093\vspace*{-0.54ex}\\
   \ttfamily\bfseries\small ameerah@alumni.caltech.edu\\[4.5ex]
\and
{Alexander Vardy}\\
   \small University of California San Diego\vspace*{-0.72ex}\\
   \small 9500 Gilman Drive, La Jolla, CA\,92093\vspace*{-0.54ex}\\
   \ttfamily\bfseries\small avardy@ucsd.edu\\[6.5ex]

\thanks{%
The research of 
Ameera Chowdhury and Alexander Vardy was supported 
by the United States National Science Foundation under 
Grants CCF-1405119 and CCF-1719139. 
This work was presented in part 
at the Allerton Conference on Communications, Control and Computing, 
October 2017, 
and at the 
IEEE International Symposium on Information Theory, June 2018.
\vspace*{-6.00ex}}
}

\newtheoremstyle{custom}
{} 
{} 
{} 
{} 
{\bfseries} 
{:} 
{.25em} 
{} 
\theoremstyle{plain}

\newtheorem{theorem}{Theorem}
\newtheorem{lemma}[theorem]{Lemma}

\newtheorem{definition}{Definition}

\newtheorem{corollary}[theorem]{Corollary}

\newtheorem{claim}{Claim}
\newtheorem{construction}{Construction}

\newtheorem*{theorem*}{Theorem}
\newtheorem*{definition*}{Definition}

\theoremstyle{definition}
\newtheorem{example}{Example}

\newcounter{enumrom}
\renewcommand{\theenumrom}{(\roman{enumrom})}

\makeatletter
\renewcommand{\@endtheorem}{\endtrivlist}
\makeatother

\makeatletter
\renewcommand{\fnum@figure}{{\bf Figure\,\@arabic\c@figure}}
\makeatother

\newcommand{\A}{{\mathcal A}}

\renewcommand{\P}{{\mathcal P}}
\newcommand{\cC}{{\mathscr{C}}}

\DeclareMathAlphabet{\mathbfsl}{OT1}{ppl}{b}{it} 

\renewcommand{\leq}{\leqslant}
\renewcommand{\ge}{\geqslant} 
\renewcommand{\geq}{\geqslant}

\newcommand\lref[1]{Lem\-ma\,\ref{lem:#1}}
\newcommand\tref[1]{The\-o\-rem\,\ref{thm:#1}}
\newcommand\cref[1]{Corollary\,\ref{cor:#1}}
\newcommand\aref[1]{Appendix\,\ref{app:#1}}
\newcommand\sref[1]{Section\,\ref{sec:#1}}
\newcommand\dref[1]{Definition\,\ref{def:#1}}

\newcommand\clref[1]{Claim\,\ref{clm:#1}}

\newcommand\eref[1]{Example\,\ref{ex:#1}}
\newcommand\cnref[1]{Con\-struct\-ion~\ref{cnstr:#1}}

\newcommand{\FF}{F}

\newcommand{\Strut}[2]{\rule[-#2]{0cm}{#1}}

\renewcommand{\epsilon}{\varepsilon}

\newcommand{\Pe}{P_{\kern-1pt\rm e}}

\makeatletter
\DeclareRobustCommand{\sbinom}{\genfrac[]\z@{}}
\makeatother

\DeclareMathOperator{\RS}{RS}
\DeclareMathOperator{\GRS}{GRS}
\DeclareMathOperator{\GF}{GF}
\newcommand{\et}{{\emph{et al.}}}
\newcommand{\BigO}[1]{\ensuremath{{O}\bigl(#1\bigr)}}

\renewcommand{\P}{{\mathbb P}}

\newcommand{\zero}{{\mathbf 0}}
\newcommand{\one}{{\mathbf 1}}

\newcommand{\Cref}[1]{Co\-ro\-lla\-ry\,\ref{#1}}

\makeatletter

\gdef\@punct{.\ \ }  
\def\@sect#1#2#3#4#5#6[#7]#8{%
  \ifnum #2>\c@secnumdepth
     \def\@svsec{}
  \else
     \refstepcounter{#1}\edef\@svsec{%
     \ifnum #2>0{{\csname the#1\endcsname}}.\fi%
    \hskip .5em}
  \fi
  \@tempskipa #5\relax
  \ifdim \@tempskipa>\z@
     \begingroup #6\relax
       \@hangfrom{\hskip #3\relax\@svsec}{\interlinepenalty \@M #8\par}
     \endgroup
     \csname #1mark\endcsname{#7}
     \addcontentsline{toc}{#1}{\ifnum #2>\c@secnumdepth\else
          \protect\numberline{\csname the#1\endcsname}\fi#7}
  \else
     \def\@svsechd{#6\hskip #3\@svsec #8\@punct\csname #1mark\endcsname{#7}
     \addcontentsline{toc}{#1}{\ifnum #2>\c@secnumdepth \else
          \protect\numberline{\csname the#1\endcsname}\fi#7}}
  \fi
  \@xsect{#5}}

\def\@ssect#1#2#3#4#5{\@tempskipa #3\relax
  \ifdim \@tempskipa>\z@
    \begingroup #4\@hangfrom{\hskip #1}{\interlinepenalty \@M #5\par}\endgroup
  \else \def\@svsechd{#4\hskip #1\relax #5\@punct}\fi
  \@xsect{#3}}

\makeatother

\begin{document}
\maketitle

\thispagestyle{empty}

\begin{abstract}
\vspace*{1.08ex}

\looseness=-1\noindent
We consider $(n,k,l)$ MDS codes of length $n$, dimension $k$, and
subpacketization $l$ over a finite field $\FF$. A~codeword of such 
a code consists 
of $n$ column-vectors of length $l$ 
over $\FF$, with the property that any $k$ of them 
suffice to recover the entire codeword. Each of these $n$ vectors may 
be stored on a separate node in a~network. If one of the $n$ nodes
fails, we can recover its content by downloading symbols from the 
surviving nodes, and the total number of symbols downloaded in the 
worst case is called the \emph{repair bandwidth} of the code. 
By the cut-set bound, the repair bandwidth 
of an $(n,k,l)$ MDS code is at least $l(n{-}1)/(n{-}k)$.\linebreak
There are several constructions of 
MDS codes whose repair
bandwidth meets or asymptotically~meets the cut-set bound. For example,
Ye and Barg constructed $(n,k,r^{n})$ Reed--Solomon 
codes that asymptotically meet the cut-set bound, 
where $r = n-k$. 
Ye and Barg also constructed optimal-bandwidth and optimal-up\-date 
$(n,k,r^{n})$ MDS codes. Wang, Tamo, and Bruck constructed
optimal-bandwidth $(n, k, r^{n/(r+1)})$ MDS codes, and these codes
have the smallest known subpacketization for optimal-bandwidth MDS
codes.

A key idea in all these constructions is to represent certain integers 
in base $r$. 
We show 
how this technique can be refined to improve the subpacketization 
of the two 
MDS code constructions by Ye and Barg, while achieving asymptotically
optimal repair bandwidth. Specifically, when $r=s^{m}$ for 
an integer~$s$,~we~ob\-tain an
$(n,k,s^{m+n-1})$ Reed--Solomon code and an optimal-update
$(n,k,s^{m+n-1})$ MDS code, both having\linebreak
asymptotically optimal repair bandwidth. 
Thus for $r = 2^m$, for example, we achieve the
subpacketization\linebreak
of $2^{m+n-1}$ rather than $2^{mn}$
in the original constructions by Ye and Barg.
When $r$ is not an integral power, we can still obtain 
$(n,k,s^{m+n-1})$ Reed--Solomon codes and optimal-update 
$(n,k,s^{m+n-1})$ MDS codes by choosing positive integers 
$s$ and $m$ such that $s^{m} \leq r$. In this case, however, 
the resulting codes have~band\-width that is near-optimal 
rather than asymptotically optimal.
We also present an extension of this idea~to 
reduce the subpacketization of the Wang--Tamo--Bruck
construction while achieving a repair-by-transfer scheme 
with asymptotically optimal repair bandwidth. 
For example, for $r = 2^m$ we achieve the 
subpacket\-ization of $2^{k/r+m-1}$, which 
significantly improves upon the 
subpacketization of $2^{mn/(r+1)}$ 
in the Wang--Tamo--Bruck construction. 
Based on the foregoing examples,
we believe our approach may be 
generally useful in reducing the subpacketization 
of MDS code constructions that utilize $r$-ary expansion.
\vspace*{6.00ex}
\end{abstract}

\newpage
\setcounter{page}{1}
\section{Introduction} 
\label{sec:Introduction}
\vspace{-1.00ex}

\noindent\looseness=-1
MDS codes are desirable for distributed storage applications because
of their optimal storage versus reliability tradeoff. Codewords of
an $(n,k,l)$ MDS code of length $n$, dimension $k$, and subpacketization $l$
over a finite field $F$ consist of $n$ column vectors of length $l$ over $F$,
with the property that any $k$ of the $n$ vectors suffice to recover the 
entire codeword.
Each of these $n$ vectors may 
be stored on a separate node in a~network.
If one of the $n$ nodes fails, a replacement node is set up to recover
the content stored at the failed node by downloading information from
the remaining functional nodes. The \textit{(exact) repair bandwidth}
is the total number of symbols downloaded in the worst case when
repairing a failed node exactly.
By the cut-set bound of \cite{Dimakis}, 
an $(n,k,l)$ MDS code has repair bandwidth at least
\begin{equation}
\label{cutset}
\left(\frac{n-1}{n-k}\right)l.
\end{equation}
Many $(n,k,l)$ MDS codes with repair bandwidth meeting or
asymptotically meeting \eqref{cutset} have been constructed; for an
excellent survey, see \cite{RTGE} and the references therein. For
example, Ye and Barg \cite{YBShort} constructed $(n,k,r^{n})$ 
Reed--Solomon codes that asymptotically meet the cut-set bound \eqref{cutset},
where $r=n-k$ is the number of parities.
Ye and Barg have also constructed in~\cite{YB}
optimal-bandwidth and optimal-update $(n,k,r^{n})$ MDS codes.
Here, \textit{optimal-bandwidth} means meeting the bound \eqref{cutset},
whereas 
\textit{optimal-update} will be defined precisely in \sref{update}.
Wang, Tamo, and Bruck~\cite{WTBExplicitPublished} constructed
optimal-bandwidth $\bigl(n,k,r^{n/(r+1)}\bigr)$ MDS codes; these~codes
have the smallest known subpacketization for optimal bandwidth MDS codes.

Optimal-bandwidth $(n,k,l)$ MDS codes necessarily have a high level of
subpacketization. For example,~a~result of \cite{AG} shows that the
subpacketization $l$ of an optimal-bandwidth $(n,k,l)$ MDS code must
satisfy
\begin{equation}
\label{lowerboundlMSR}
l \,\geq\, \exp\left(\frac{k-1}{4r}\,\right).
\end{equation}
The high-rate optimal-bandwidth codes referenced in \cite{RTGE}
typically require an even larger subpacketization. The bound \eqref{lowerboundlMSR} is a recent improvement on a result of \cite{GTC} that showed optimal-bandwidth $(n,k,l)$ MDS codes must have subpacketization at least $l \approx \exp( \sqrt{k/r} )$.
For various reasons, which we do not discuss
here, it is often desirable to design repair schemes that achieve 
low repair bandwidth without requiring a very high 
subpacketization. Consequently this paper, 
like \cite{AV,AV2,GJ,GR,RTGE,Zigzag}, 
explores a tradeoff between the subpacketization $l$ and
the repair bandwidth of MDS codes.

\begin{table*}[ht]
\centering
\vspace{1.80ex}
\captionsetup{justification=centering}
\begin{center}{\normalsize \begin{tabular}{| c | c | c | c |} 
\hline 
\bf Code Construction & 
\bf Repair Bandwidth & 
\bf Subpacketization & 
Meets Cut-Set Bound \eqref{cutset}? 
\\ 
\hline
\begin{tabular}[c]{@{}c@{}} 
$(n,k)$ RS code \\[-0.36ex]
our construction
\end{tabular} & 
\Strut{3.6ex}{0ex}$<\frac{(n-1+3r)l}{r}$ ~for $r=s^{m}$ & $l = s^{m+n-1}$ & 
Asymptotically meets \eqref{cutset} for $r=s^{m}$ \\ 
\hline
\begin{tabular}[c]{@{}c@{}} $(n,k)$ RS code \cite{YBShort}
\end{tabular} & 
\Strut{5.2ex}{1.8ex}$<\frac{(n+1)l}{r}$ & 
$l=r^n$ & Asymptotically meets \eqref{cutset}  
\\  
\hline
\end{tabular} 
\caption{\small
Tradeoff between repair bandwidth and subpacketization for Reed--Solomon code constructions in \cite{YBShort} and~in~\sref{main} (\tref{main}) of this paper. There are no requirements on the field size for either of the constructions.} 
\label{tab:RStradeoff}}
\vspace{-1.80ex}
\end{center}
\end{table*}

\begin{table*}[ht]
\centering
\captionsetup{justification=centering}
\begin{center}{\normalsize 
\begin{tabular}{| c | c | c | c |} 
\hline
\bf Code Construction & 
\bf Repair Bandwidth  & 
\bf Subpacketization  & 
\begin{tabular}[c]{@{}c@{}}
 Optimal Update\\[-.36ex]
 with Diagonal\\[-.36ex]
 Encoding Matrices 
\end{tabular}
\\ \hline
\begin{tabular}[c]{@{}c@{}} $(n,k,l)$ MDS code \\[-0.36ex]
our construction
\end{tabular} &
\begin{tabular}[c]{@{}c@{}} 
\Strut{6.0ex}{1.80ex}$<\frac{\left(n-1 + 2 \sum_{v=1}^{m-1} (s^{v}-1)\right)l}{r}$ ~for $r=s^{m}$\\
asymptotically meets \eqref{cutset}\Strut{1.0ex}{2.0ex}
\end{tabular} & $l = s^{m+n-1}$ & Yes  
\\ 
\hline
\begin{tabular}[c]{@{}c@{}} $(n,k,l)$ MDS code  \cite{YB}
\end{tabular} & \Strut{5.2ex}{1.8ex} $\frac{(n-1)l}{r}$, ~meets \eqref{cutset}  & $l=r^n$ & Yes  \\  \hline
\end{tabular} 
\caption{\small
Tradeoff between repair bandwidth and subpacketization 
for optimal-update $(n,k,l)$ MDS code constructions in \cite{YB}
and in \sref{update} (\cnref{adapt} and \tref{bb}) of this paper. Both constructions require $|F| \geq rn$.} 
\label{tab:OUtradeoff}}
\end{center}
\vspace*{-1.80ex}
\end{table*}

\looseness=-1
A key idea in many $(n,k,l)$ MDS code constructions
\cite{WTBExplicitPublished,YB,YBShort,RTGE,Zigzag,CHLPerm} is to
expand integers in base $r$. In this~paper, we demonstrate how this
technique can be refined to improve the subpacketization of the two
$(n,k,l)$ MDS code constructions by Ye and Barg \cite{YB,YBShort}. 
Specifically, in the case where the number of parities $r$ is 
an integral power, that is $r=s^{m}$, we construct an
$(n,k,s^{m+n-1})$ Reed--Solomon code and an optimal-update
$(n,k,s^{m+n-1})$ MDS code, both having 
an asymptotically optimal repair bandwidth. 
For example for $r = 2^m$, we achieve the
subpacketization of $2^{m+n-1}$, which improves upon the
subpacketization of $2^{mn}$ in the constructions of Ye and Barg. 
Table\,\ref{tab:RStradeoff} compares our Reed--Solomon code
construction with that of Ye and Barg \cite{YBShort}. \tref{main}
in \sref{main} gives a precise statement of our main result for Reed--Solomon
codes. Table\,\ref{tab:OUtradeoff} compares our optimal-update MDS
code construction with that of Ye and Barg \cite{YB}. \cnref{adapt}
and \tref{bb} in \sref{update} give a precise statement of our main result for 
optimal-update MDS codes.

When $r$ is not an integral power, we can still obtain
$(n,k,s^{m+n-1})$ Reed--Solomon codes and optimal-update
$(n,k,s^{m+n-1})$ MDS codes by choosing positive integers $s$ and $m$
such that $s^{m} \leq r$. In this case, however, the resulting codes
have bandwidth that is near-optimal rather than asymptotically optimal.

We also present another refinement of the $r$-ary expansion technique that
leads to $(n,k,l)$ MDS codes with repair-by-transfer schemes that have
asymptotically optimal repair bandwidth.
In a \textit{repair-by-transfer scheme},
exact repair of 
a failed node involves downloading from the surviving nodes only 
certain subsets of their symbols rather than linear combinations thereof.
Thus the contacted nodes have to perform minimal
computations, which makes repair-by-transfer schemes an attractive
variant of linear repair schemes.

\looseness=-1
Specifically, 
when $r$ is an integral power, we significantly reduce 
the subpacketization of the Wang--Tamo--Bruck construction
\cite{WTBExplicitPublished} while achieving a repair-by-transfer
scheme that has asymptotically optimal~repair~band\-width. 
For example for $r = 2^m$, we achieve the subpacketization of
$2^{k/r+m-1}$, which improves upon 
the subpacketization of
$2^{mn/(r+1)}$ in the Wang--Tamo--Bruck construction~\cite{WTBExplicitPublished}.
In general, when
$r = s^m$~for~an~integer $s \ge 2$, 
our codes have subpacketization $l = s^{k/r+m-1}\kern-1pt =
rs^{k/r-1}$. Specifically, 
we construct an $(n,k,s^{k/r+m-1})$ MDS code which has a repair-by-transfer scheme
with asymptotically optimal repair bandwidth. \cnref{wtbadapt} and
\tref{wtbadaptbw} in \sref{ourcode} 
give a precise statement of our main result for
repair-by-transfer schemes.

Table\,\ref{tab:tradeoff} compares our code construction with other
state-of-the-art $(n,k,l)$ MDS codes. 
Rawat, Tamo, Guruswami, and Efremenko recently proposed a code construction 
in \cite[Theorem III.1]{RTGE} which utilizes a design parameter $\tau$ to
obtain a family of MDS codes with small subpacketization and
near-optimal bandwidth. When $r=s^{m}$ is an integral power, 
setting $\tau = \lceil \frac{n}{mr} \rceil$ 
in \cite[Theorem III.1]{RTGE} yields an $(n,k,l)$ MDS code 
with~asymptotically optimal bandwidth that performs as well
as the codes we construct in \sref{ourcode}. 
We point out that our construction is contemporaneous
with and very different from the construction in \cite[TheoremIII.1]{RTGE}.

\begin{table*}[ht]
\centering
\captionsetup{justification=centering}
\begin{center}{\small \begin{tabular}{| c | c | c |@{}c@{}| c |} 
\hline 
\begin{tabular}[c]{@{}c@{}} \bf Code\\[-0.36ex] \bf Construction \end{tabular} &
\bf Subpacketization & 
\bf Repair Bandwidth & 
\bf Field Size & 
\begin{tabular}[c]{@{}c@{}} Repair by \\transfer? \end{tabular} 
\\ 
\hline
\begin{tabular}[c]{@{}c@{}} Ye--Barg \cite{YB} \end{tabular} & 
\begin{tabular}[c]{@{}c@{}} $l=r^{\left \lceil \frac{n}{r} \right \rceil}$ \end{tabular} & \begin{tabular}[c]{@{}c@{}} $\frac{(n-1)l}{r}$, ~meets \eqref{cutset}  \end{tabular} & \begin{tabular}[c]{@{}c@{}} \Strut{3.9ex}{0.90ex}$< r \left \lceil \frac{n}{r} \right \rceil$\\ fully explicit\\[0.54ex] \end{tabular} & Yes \\ \hline
\begin{tabular}[c]{@{}c@{}} Wang-Tamo-Bruck\cite{WTBExplicitPublished}
\end{tabular} & $l=r^{n/(r+1)}$ & \begin{tabular}[c]{@{}c@{}} $\frac{(n-1)l}{r}$, ~meets \eqref{cutset}  \end{tabular}  & \begin{tabular}[c]{@{}c@{}} \Strut{3.9ex}{0.90ex}$< \binom{n-1}{r}rl$\\ not fully explicit \\[0.54ex] \end{tabular}  & No \\   \hline
\begin{tabular}[c]{@{}c@{}} this paper
\end{tabular} & \begin{tabular}[c]{@{}c@{}} $l = s^{(k/r)+m-1}$\\ for $r = s^m$ \end{tabular} & \begin{tabular}[c]{@{}c@{}} $<\left(1 + \frac{2\! \sum\limits_{v=1}^{m-1} (r-s^{v})}{n-1} \right)\Strut{6.84ex}{0ex} \frac{(n-1)l}{r}$\\[0.72ex] Asymptotically meets \eqref{cutset}\Strut{2.00ex}{1.80ex} \end{tabular} & \begin{tabular}[c]{@{}c@{}} $< \binom{n-1}{r}rl$\\[0.72ex] not fully explicit \end{tabular} & Yes \\ 
\hline
\begin{tabular}[c]{@{}c@{}} Rawat \et \cite{RTGE}\\ (special case) 
\end{tabular} & \begin{tabular}[c]{@{}c@{}} $l = r^{\left \lceil \frac{n}{Cr} \right \rceil}$\\[1.08ex] for $C \geq 1$ \end{tabular} & \begin{tabular}[c]{@{}c@{}} $<\left(1 + \frac{1}{\left \lceil \frac{n}{Cr} \right \rceil}\right) \frac{(n-1)l}{r}$\\[1.80ex] Asymptotically meets \eqref{cutset} \end{tabular} & 
\begin{tabular}[c]{@{}c@{}} 
~~\Strut{15.66ex}{7.38ex}$\begin{cases} \geq (n+1)^{(r-1)l+1},\\[-0.54ex] 
\hfill\mbox{fully explicit}\hspace*{-1.62ex}\\[0.90ex]
< \BigO{n^{r}rl},\\[-0.54ex] 
\hfill\mbox{not fully explicit}\hspace*{-1.62ex} \end{cases}$ 
\end{tabular} & 
Yes \\  \hline
\end{tabular} 
\caption{\small
Comparisons of repair bandwidth, subpacketization, field size, and existence of repair-by-transfer schemes for the code proposed in \sref{ourcode} (\cnref{wtbadapt} and \tref{wtbadaptbw}) and other state-of-the-art $(n,k,l)$ MDS codes.}
\label{tab:tradeoff}}
\end{center}
\end{table*}

Having demonstrated how to improve the subpacketization of three
very different $(n,k,l)$ MDS code constructions, we hope that our
approach may be broadly useful in reducing the subpacketization of
MDS codes. 
The general idea is simple: when $r=s^{m}$ is an integral power, 
one can improve the subpacketization of MDS code 
constructions that utilize $r$-ary expansion 
by instead expanding integers in the smaller base $s$. 
The results presented in this paper 
show how to tailor
this general idea to specific $(n,k,l)$ MDS code constructions.

The rest of this paper is organized as follows.
We begin in \sref{rs} with some background and context for the repair of
Reed--Solomon codes. \sref{technicalbg} briefly discusses the
Guruswami--Wootters~\cite{GW} characterization of linear exact repair schemes
for MDS codes, upon which our construction as well as the 
constructions in \cite{GW, TYB, YBShort, DDKM1, DDKM2, DD, DM, BW} are
based. \sref{main} presents our main result for Reed--Solomon codes
(\tref{main}). \sref{update} presents 
\cnref{adapt} and \tref{bb}, which constitute our main results
for optimal-update MDS codes. 
\sref{framework} adapts 
the Wang--Tamo--Bruck framework~\cite{WTBExplicitPublished}
for use
in constructing $(n,k,l)$ MDS codes that have repair-by-transfer
schemes with asymptotically optimal repair bandwidth. \sref{ourcode}
presents our main results (\cnref{wtbadapt} and \tref{wtbadaptbw}) on
repair-by-transfer schemes with asymptotically optimal repair
bandwidth.

$\,$\vspace{1.80ex}
\section{Repairing Reed--Solomon Codes}
\label{sec:rs}

In this section, we first provide some background and context for the repair 
of Reed--Solomon codes. Subsequently, we discuss 
the Guruswami--Wootters~\cite{GW} characterization 
of linear exact repair schemes for MDS codes,
upon which our result as well as the results of
\cite{GW, TYB, YBShort, DDKM1, DDKM2, DD, DM, BW} are based.

\subsection{General overview}
\label{sec:rsoverview}
In the conventional solution to the exact repair problem using
{Reed--Solomon} (RS) codes, we split the file into $k$ blocks. Each of
the $k$ blocks is represented by some element of a finite field $E$
and then viewed as the coefficient of a polynomial. In this way, the
file is identified with a polynomial $f$ over $E$ of degree
$\leq k-1$. We then distribute the file over $n$ nodes by choosing $n$
distinct evaluation points $\alpha_{1}, \ldots, \alpha_{n} \in E$ and
storing $f(\alpha_{i})$ at node $i$. To recover a failed node, we can
download information from any $k$ remaining nodes because any $k$
evaluations $f(\alpha_{i})$ of a degree $k-1$ polynomial exactly
determine the polynomial and hence the contents of the failed node.
One can show that, in the conventional solution, downloading
information from $k$ remaining nodes is not only sufficient for
repairing one node, but also necessary. Therefore, at first glance, 
RS codes seem ill-suited for the exact repair problem because recovering 
the contents of a single failed node requires downloading $k$ symbols of
$E$ or, equivalently, the whole file. Thus, despite the ubiquity of RS
codes in storage systems, until the recent work of Guruswami and
Wootters \cite{GW}, these codes were regarded as poorly suited for
distributed storage~applications since they were thought to have a
very high repair bandwidth.

To mitigate this issue, we can apply the \textit{regenerating codes}
framework \cite{Dimakis, SPDC} in which a replacement node may
download only \textit{part} of the contents of a surviving node rather
than being forced to download the whole node. This is accomplished by
viewing $E$ as a vector space over one of its subfields $F$, and
allowing each surviving node to return one or more symbols in~$F$. 
Crucially, each node may return fewer than $\log_{|F|}|E|$
symbols of $F$ when queried, and our goal is to download as few such
subsymbols as possible. The (exact) \textit{repair bandwidth} of the
code over $F$ is the total number of subsymbols downloaded in the
worst case. We assume that each node returns an $F$-linear function of
its contents so that we have a \textit{linear repair scheme}.

Since Reed--Solomon codes are MDS, the cut-set bound \eqref{cutset}
and the lower bound \eqref{lowerboundlMSR} from \cite{GTC}
apply. Recently, Tamo, Ye, and Barg \cite{TYB, TYB2} improved
\eqref{lowerboundlMSR} for RS codes, and showed that any RS code
meeting the cut-set bound has a subpacketization $l$ that satisfies
\begin{equation}
\label{TYB-lb}
l \,\geq\, \exp\Bigl(\bigl(1+o(1)\bigr)k \log k\Bigr).
\end{equation}
They also explicitly constructed in~\cite{TYB, TYB2} Reed--Solomon codes 
meeting the cut-set bound whose subpacketization is given by
\begin{equation*}
l \,=\, \exp\Bigl(\bigl(1+o(1)\bigr)n \log n\Bigr).
\end{equation*}
We note that the lower bound on the subpacketization 
in \eqref{TYB-lb} does not apply to RS codes whose repair
bandwidth meets the cut-set bound only \emph{asymptotically} as
$n \to \infty$.
For example, Ye and Barg~\cite{YBShort} have previously constructed 
such RS codes (whose repair bandwidth asymptotically 
meets the 
cut-set bound) with subpacketization $l=r^{n}$.
If $r$ is fixed, while $n \to \infty$, then $r^n$ could 
be significantly lower than 
\eqref{TYB-lb}.

When the subpacketization $l$ is small, we cannot hope to meet the
cut-set bound. In particular, Guruswami and Wootters \cite{GW} established
a lower bound on the repair bandwidth of $(n,k,l)$ MDS codes with 
linear repair schemes. Subsequently, Dau and Milenkovic \cite{DM} 
refined and improved this bound in some cases. As a result, it is
now known that an $(n,k,l)$ MDS code with a linear repair scheme
must have bandwidth at least:
\begin{equation}
\label{GWlower}
(n-1) \log_{|F|}\left(\frac{(n-1)|E|}{(r-1)(|E|-1)+(n-1)}\right).
\end{equation}
Moreover, full-length RS codes meeting \eqref{GWlower} were
explicitly constructed in \cite{GW} and \cite{DM}. Necessarily, these
codes have small subpacketization $l$; the optimal-bandwidth RS code
in \cite{GW} has subpacketization $l = \log_{n/r}(n)$, for example. In
Table\,\ref{tab:rstradeoff}, we summarize the tradeoffs between
subpacketization and repair bandwidth for explicit RS code
constructions. Further work on repairing RS codes can be found in
\cite{DDKM1, DDKM2, DD, BW, LWJ}.

\begin{table*}[ht]
\centering
\captionsetup{justification=centering}
\begin{center}{\normalsize \begin{tabular}{| c | c | c | c |} 
\hline 
\bf Code Construction & 
\bf Repair Bandwidth &
\bf Subpacketization  & Meets Cut-Set Bound \eqref{cutset}? \\ \hline
\begin{tabular}[c]{@{}c@{}} $(n,k)$ RS code  \cite{GW}
\end{tabular} & $n-1$  & \Strut{4.50ex}{1.80ex}$l=\log_{n/r} n$ & No, but meets \eqref{GWlower}  \\   \hline
\begin{tabular}[c]{@{}c@{}} $(n,k)$ RS code  \cite{DM}\\  
\end{tabular} &  $(n-1)l(1-\log_nr)$ & \Strut{4.50ex}{1.80ex}$\log_q n$ & No, but meets \eqref{GWlower}  \\   \hline
\begin{tabular}[c]{@{}c@{}} $(n,k)$ RS code\\[-0.54ex]  this paper
\end{tabular} & $<\frac{(n-1+3r)l}{r}$ ~for $r=s^{m}$ & $l = s^{m+n-1}$ & Asymptotically meets \eqref{cutset} for $r=s^{m}$  \\ 
\hline
\begin{tabular}[c]{@{}c@{}} $(n,k)$ RS code  \cite{YBShort}
\end{tabular} & \Strut{5.40ex}{2.16ex} $<\frac{(n+1)l}{r}$  & $l=r^n$ & Asymptotically meets \eqref{cutset}  \\  \hline
\begin{tabular}[c]{@{}c@{}} $(n,k)$ RS code \cite{TYB}
\end{tabular} & \Strut{5.40ex}{2.16ex}$\frac{(n-1)l}{r}$  & $l\approx n^{ n}$ & Yes, meets \eqref{cutset}  \\ 
\hline
\end{tabular} 
\caption{Tradeoff between repair bandwidth and subpacketization for Reed--Solomon codes. There are no requirements on the size of the finite field for any of the constructions.} 
\label{tab:rstradeoff}}
\end{center}
\end{table*}


\subsection{Linear repair schemes for Reed--Solomon codes}
\label{sec:technicalbg}

In a Reed--Solomon code, a codeword is a sequence of function values of 
a~polynomial of degree less than $k$. Given a finite field $E$, let
$E[x]$ denote the ring of polynomials over $E$.

\begin{definition}
\label{def:RS}
A generalized Reed--Solomon code, denoted $\GRS(n,k,A,\nu)$, of dimension $k$ over the finite field $E$ using $n$ evaluation points $A = \{\alpha_{1}, \ldots, \alpha_{n}\} \subset E$ is the set of vectors
\[
\Bigl\{(\nu_{1}f(\alpha_{1}), \ldots, \nu_{n}f(\alpha_{n})) : f \in E[x], \deg(f) < k \Bigr\} \subseteq E^{n},
\]
where $\nu = (\nu_{1}, \ldots, \nu_{n})$ are some nonzero coefficients
in $E$. When $\nu = (1, \ldots, 1)$, the corresponding generalized Reed--Solomon code $\GRS(n,k,A,\one) = \RS(n,k,A)$ is called a Reed--Solomon code. 
\end{definition}

By \cite[Theorem 4 in Chapter 10]{MWS},
we have that $\RS(n,k,A)^{\perp}$, the dual of a Reed-Solomon code $\RS(n,k,A)$, is a generalized Reed-Solomon code $\GRS(n,n-k,A,\nu)$, where 
\begin{equation}
\label{nuvec}
\nu_{i} = \prod_{j \neq i} (\alpha_{i} - \alpha_{j})^{-1}
\hspace{6ex}
\text{for $i = 1,2,\ldots,n$}
\end{equation}

We now formalize the definition of a linear repair scheme for the
Reed-Solomon code $\RS(n,k,A)$ discussed above. Recall that
each node $\alpha$ returns an $F$-linear function of its contents
$f(\alpha)$ where $F$ is a subfield of $E$. One example of an
$F$-linear function from $E$ to $F$ is the \textit{field trace}
$\mathrm{tr}_{E/F}$.
\begin{definition}
\label{def:fieldtrace}
Let $E = \GF\bigl(q^l\bigr)$ be an extension of degree $l$ of 
the field $F = \GF(q)$.
The field trace 
\mbox{$\mathrm{tr}_{E/F}\!: E \rightarrow F$} is defined to be
\[
\mathrm{tr}_{E/F}(\beta) = \beta + \beta^{q} + \beta^{q^2} + \cdots + \beta^{q^{l-1}}.
\]
\end{definition}

\noindent 
Conversely, one can show that the $F$-linear functions from $E$ to $F$
are precisely the \textit{trace functionals} 
\mbox{$L_{\gamma}:E \rightarrow F$} 
given by $L_{\gamma}(\beta) = \mathrm{tr}_{E/F}(\gamma \beta)$ 
for some $\gamma \in E$.

In a linear repair scheme, a node that stores $f(\alpha)$ therefore
returns elements of $F$ of the form $L_{\gamma}(f(\alpha))$. The field
elements $\gamma \in E$ used by each node thus describe a linear
repair scheme for $\RS(n,k,A)$. The following definition of a linear
exact repair scheme is from Guruswami and Wootters~\cite{GW}.

\begin{definition}
\label{def:repairscheme}
A \textit{linear exact repair scheme} for $\RS(n,k,A)$ 
over a~sub\-field $F \subseteq E$ consists of
\begin{itemize}
 \item For each $\alpha_{i} \in A$ and for each $\alpha_{j} \in A \setminus \{\alpha_{i}\}$, a set of queries $Q_{j}(i) \subseteq E$.
 \item For each $\alpha_{i} \in A$, a linear reconstruction algorithm that computes 
 \begin{equation}
 \label{coeffs}
 f(\alpha_{i}) = \sum_{h=1}^{l} \lambda_{h} \mu_{h}
 \end{equation}
for some coefficients $\lambda_{h} \in F$ and a basis 
$\mu_{1}, \ldots, \mu_{l}$ for $E$ over $F$, where
the coefficients $\lambda_{h}$ are $F$-linear combinations 
of the responses to the queries, namely:
 \begin{equation}
 \label{queries}
 \bigcup_{\hspace*{-1.80ex}j \in [n] \setminus \{i\}} 
\!\Bigl\{L_{\gamma}\bigl(f(\alpha_{j})\bigr) : \gamma \in Q_{j}(i) \Bigr\}.
 \end{equation}
\end{itemize}
\noindent 
The \textit{repair bandwidth} $b$ of the linear exact repair scheme is
the total number of symbols in $F$ downloaded to recover a failed node
in the worst case, which can be expressed as:
\begin{equation}
\label{bw}
b \,=\, \max_{i \in [n]} \sum_{j \in [n] \setminus \{i\}} \bigl|Q_{j}(i)\bigr|.
\end{equation}
\end{definition}

Recall that $l = \log_{|F|}|E|$ is the dimension of $E$ as a vector
space over $F$. Guruswami and Wootters \cite{GW} show that specifying
a linear repair scheme for $\RS(n,k,A)$ over $F$ is equivalent to
finding, for each $\alpha_{i}$ in $A$, a~set of $l$ polynomials
$\P_{i} \subset E[x]$ of degree less than $n-k$ such that
$\{p(\alpha_{i}) : p \in \P_{i}\}$ is a basis for $E$~over~$F$
(whereas $\{p(\alpha_{j}) : p \,{\in}\, \P_{i}\}$ for all $j \neq i$
spans a low-dimensional subspace over $F$).
Specifically, the following theorem is due to 
Guruswami and Wootters~\cite{GW}.

\begin{theorem}
\label{thm:characterization}
Let $F \subseteq E$ be a subfield such that the degree of $E$ over $F$
is $l$ and let $A \subset E$ be any set of evaluation points. The
following are equivalent.
\begin{enumerate}
 \item There is a linear repair scheme for $\RS(n,k,A)$ over $F$ with bandwidth $b$.
 \item For each $\alpha_{i} \in A$, there is a set $\P_{i} \subset E[x]$ of\, $l$ polynomials of degree less than $n-k$ such that
 \begin{equation}
 \label{basiscondition}
 \dim_{F}\Bigl(\{p(\alpha_{i}) : p \in \P_{i}\}\Bigr) = \,l
 \end{equation}
 and the sets $\{p(\alpha_{j}): p \in \P_{i} \}$ for $j \neq i$ satisfy
 \begin{equation}
 \label{smallbandwidth}
 b ~\geq\!\! \sum_{j \in [n] \setminus \{i\}} 
   \!\!\!\dim_{F}\Bigr(\{p(\alpha_{j}): p \in \P_{i}\}\Bigl). 
 \end{equation}
\end{enumerate}
\end{theorem}
\noindent The RS code constructed in \sref{main}, as well as the RS codes constructed in \cite{TYB, YBShort, GW, DDKM1, DDKM2, DD, DM, BW}, rely on the fact that the second statement in \tref{characterization} implies the first, so we sketch the proof.

Suppose that the codeword symbol $f(\alpha_{i})$ in a codeword
$(f(\alpha_{1}), \ldots, f(\alpha_{n})) \in \RS(n,k,A)$ is erased.
\begin{lemma}
\label{lem:dualbasis}
For a basis $\{\zeta_{1}, \ldots, \zeta_{l}\}$ for $E$ over $F$, the value of $f(\alpha_{i})$ can be uniquely recovered from the values 
\[
\{\mathrm{tr}_{E/F}(\zeta_{h} f(\alpha_{i}))\}_{h=1}^{l}.
\]
\end{lemma}

\noindent \textbf{Proof.} 
If $\{\mu_{1}, \ldots, \mu_{l}\}$ is the dual (trace-orthogonal) basis of the basis $\{\zeta_{1}, \ldots, \zeta_{l}\}$, then
\begin{equation}
\label{coeffsinbasis}
f(\alpha_{i}) = \sum_{h=1}^{l} \mathrm{tr}_{E/F}(\zeta_{h}f(\alpha_{i})) \mu_{h}. \qed
\end{equation}

Suppose we have $l$ codewords $\{c_{h}^{\perp} = (c_{h,1}^{\perp}, \ldots, c_{h,n}^{\perp})\}_{h=1}^{l}$ in $\RS(n,k,A)^{\perp}$ such that $\{c_{1,i}^{\perp}, \ldots, c_{l,i}^{\perp}\}$ is a basis for $E$ over $F$. By \lref{dualbasis}, to find the value of $f(\alpha_{i})$, it suffices to find the values of $\{\mathrm{tr}_{E/F}(c_{h,i}^{\perp}f(\alpha_{i}))\}_{h=1}^{l}$.

\begin{lemma}
\label{lem:duality}
From $\{\{\mathrm{tr}_{E/F}(c_{h,j}^{\perp}f(\alpha_{j}))\}_{h=1}^{l}\}_{j \in [n]\setminus \{i\}}$, we can recover the values of $\{\mathrm{tr}_{E/F}(c_{h,i}^{\perp}f(\alpha_{i}))\}_{h=1}^{l}$. 
\end{lemma}

\noindent \textbf{Proof.} 
By duality and because $\mathrm{tr}_{E/F}$ is $F$-linear, we have for all $h \in [l]$ that
\begin{equation*}
\mathrm{tr}_{E/F}(c_{h,i}^{\perp}f(\alpha_{i})) = -\sum_{j \neq i} \mathrm{tr}_{E/F}(c_{h,j}^{\perp}f(\alpha_{j})). \qed
\end{equation*}

Similarly, \lref{querytime} holds because $\mathrm{tr}_{E/F}$ is $F$-linear.
\begin{lemma}
\label{lem:querytime}
Define $Q_{j}(i)$ to be a maximum linearly independent subset of the set $\{c_{h,j}^{\perp}\}_{h=1}^{l}$. We can find the values of 
\[
\{\mathrm{tr}_{E/F}(c_{h,j}^{\perp}f(\alpha_{j}))\}_{h=1}^{l}
\]
for $j \in [n] \setminus \{i\}$ from the values of 
\begin{equation*}
\{\mathrm{tr}_{E/F}(\gamma f(\alpha_{j})) : \gamma \in Q_{j}(i) \} = \{L_{\gamma}(f(\alpha_{j})) : \gamma \in Q_{j}(i) \}.
\end{equation*}
\end{lemma}

Now we show how to specify a linear exact repair scheme for $\RS(n,k,A)$ over $F$ as in \dref{repairscheme}.
\begin{lemma}
\label{lem:dualtorepair}
A set of\, $l$ codewords $\{c_{h}^{\perp} = (c_{h,1}^{\perp}, \ldots, c_{h,n}^{\perp})\}_{h=1}^{l}$ in $\RS(n,k,A)^{\perp}$ such that $\{c_{1,i}^{\perp}, \ldots, c_{l,i}^{\perp}\}$ is a basis for $E$ over $F$ suffices to specify a linear exact repair scheme for $\RS(n,k,A)$ over $F$ as in \dref{repairscheme}.
\end{lemma}

\noindent \textbf{Proof.} 
Setting $\lambda_{h} = \mathrm{tr}_{E/F}(c_{h,i}^{\perp}f(\alpha_{i}))$ and letting $\mu_{1}, \ldots, \mu_{l}$ be a dual basis for the basis $\{c_{1,i}^{\perp}, \ldots, c_{l,i}^{\perp}\}$, we have that \eqref{coeffs} holds by \eqref{coeffsinbasis}. Moreover, defining $Q_{j}(i)$ as in \lref{querytime}, we see that the coefficients $\lambda_{h}$ are $F$-linear combinations of the queries in \eqref{queries} by \lref{duality} and \lref{querytime}. \qed

\vspace{0.25cm}
Since $\RS(n,k,A)^{\perp} = \GRS(n,n-k,A,\nu)$, where $\nu$ is given by \eqref{nuvec}, we see that the hypothesis of \lref{dualtorepair} is equivalent to \eqref{basiscondition} and that the right hand sides of \eqref{bw} and \eqref{smallbandwidth} are equal. Hence, the second statement of \tref{characterization} implies the first. 

\vspace{3.0ex}

\section{Reed--Solomon Repair Schemes with Improved Subpacketization}
\label{sec:main}

\noindent
Recall that Ye and Barg \cite{YBShort} explicitly constructed
$(n,k,r^{n})$ RS codes whose bandwidth asymptotically meets the
cut-set bound. In this section, we generalize this construction and
improve the subpacketization at the expense of an asymptotically negligible
increase in repair bandwidth. Table\,\ref{tab:RStradeoff} compares 
our Reed--Solomon code construction with that of Ye and Barg \cite{YBShort}. 
If $r = 2^m$, for example, we achieve
the subpacketization of~$2^{m+n-1}$,\linebreak
which improves upon the
subpacketization of $2^{mn}$ in the constructions by Ye and Barg.
\tref{main} gives~a~precise statement of our main result for RS codes.

\begin{theorem}
\label{thm:main}
\looseness=-1
Let $n$ and $k$ be arbitrary fixed integers, and suppose that $n-k = s^{m}$ 
where $s \geq 2$ and $m \geq 1$. Let $F$ be a finite field and
let $h(x)$ be a degree $l$ irreducible polynomial over $F$ where $l =
s^{m+n-1}$. Let $\beta$ be a root of $h(x)$ and set the symbol field
$E = F(\beta)$ to be the field generated by $\beta$ over $F$. Choose
the set of evaluation points to be 
$A = \bigl\{\beta^{s^{0}},\beta^{s^{1}}, \ldots, \beta^{s^{n-1}}\bigr\}$. 
The exact repair bandwidth
of the code $\RS(n,k,A)$ over $F$ is at most
\begin{equation}
\label{bwbound}
\left(\frac{n-1+3s^{m-1} + 2s^{m-2} + \cdots + 2s - (m-4)}{n-k}\right)l,
\end{equation}
and hence asymptotically meets \eqref{cutset} for fixed $n-k$ as $n \to \infty$.
\end{theorem}

Note that the construction in \tref{main} generalizes the Ye--Barg
construction in \cite{YBShort} because setting $s = r$ and $m=1$ in
\tref{main} yields their result.
However, in general when $r=s^{m}$ for $m \ge 2$, instead of expanding
integers in base $r$ as Ye and Barg \cite{YBShort} do, we will expand
integers in base $s$.

When $r$ is not an integral power, we can still use the ideas in the
proof of \tref{main}. For example, we can choose any positive integers
$s$ and $m$ such that $s^{m} \leq r$. The statement and proof of
\tref{main} still hold if we replace by $s^{m}$ every occurrence of
$n-k$. For fixed $r$, as $n \rightarrow \infty$, the ratio between the
repair bandwidth of the resulting RS codes and \eqref{cutset} would 
be $r/s^{m}$.
It is always possible to choose $s^{m} \leq r$ so that $r/s^{m}$
is at most $2$. We summarize this result in the
following corollary.

\begin{corollary}
\label{cor:nopower}
Let $n$ and $k$ be arbitrary fixed integers and suppose that $n-k \geq s^{m}$ where $s \geq 2$ and $m \geq 1$. The exact repair bandwidth over $F$ of the code $\RS(n,k,A)$ constructed in \tref{main} is at most
\begin{equation}
\label{bwboundnopower}
\left(\frac{n-1+3s^{m-1} + 2s^{m-2} + \cdots + 2s - (m-4)}{s^m}\right)l.
\end{equation}
For fixed $n-k$, as $n \rightarrow \infty$, the ratio between \eqref{bwboundnopower} and \eqref{cutset} would be $(n-k)/s^{m}$, which is at most $2$.
\end{corollary}

In this section, we prove \tref{weakmain}, a version of \tref{main}
with a slightly weaker bound on the repair~bandwidth.
More involved counting and case analysis, which we defer
to \aref{appendixa}, yields the repair bandwidth
bound \eqref{bwbound} in \tref{main}.
We use the notation $[x,y] = \{x, x+1, \ldots, y\}$
for integers $x<y$.

\begin{theorem}
\label{thm:weakmain}
The exact repair bandwidth over F of the code $\RS(n,k,A)$ constructed in \tref{main} is at most
\begin{equation}
\label{weakbwbound}
\left(\frac{n-1 + 2(m-1)s^m -2m +6}{n-k}\right)l,
\end{equation}
and hence asymptotically meets \eqref{cutset} for fixed $n-k$ as $n \to \infty$.
\end{theorem}

\vspace{0.25cm}

\noindent \textbf{Proof of \tref{weakmain}.} 
By \tref{characterization}, it suffices to find, for each $i \in
[0,n-1]$, a set of $l$ polynomials~$\{f_{i,j}\}_{j=1}^{l}$ satisfying
$\deg(f_{i,j}) < n-k$ so that 
\smash{$f_{i,1}(\beta^{s^{i}}), \ldots, f_{i,l}(\beta^{s^{i}})$}
form a basis for $E$ over $F$ and so that
\begin{equation}
\label{sumofdimensions}
\sum_{\substack{0 \leq t \leq n-1 \\ t \neq i}} \dim_{F} \left( \{f_{i,j}(\beta^{s^{t}})\}_{j=1}^{l} \right) 
\end{equation}
is bounded above by \eqref{weakbwbound}.
Recall that $l = s^{m+n-1}$ by assumption.
Given $a \in [0,l-1]$, we can write 
its $s$-ary expansion as $(a_{m+n-2}, \ldots, a_{0})$; that is
\begin{equation*}
a ~=\! \sum_{j=0}^{m+n-2} a_{j}s^{j},
\end{equation*}

\begin{figure*}[t]
\centering
\captionsetup{justification=centering}
\begin{center}
\begin{tikzpicture}[thick]
  \node[rectangle,draw,label=left:${a}~$,label=above:$n+1$](a){*};
  \node[inner sep=0,minimum size=0,right of=a] (b) {$\cdots$}; 
  \node[draw,rectangle,right of=b, label=above:$i+3$] (c) {*};
  \node[rectangle,draw,label=above:$i+2$,right of =c](d){0};
  \node[draw,rectangle,label=above:$i+1$,right of=d] (e) {0};
  \node[draw,rectangle,label=above:$i$,right of=e] (f) {0};
  \node[draw,rectangle,label=above:$i-1$,right of=f] (g) {*};
  \node[draw,rectangle,label=above:$i-2$,right of=g] (h) {*};
  \node[draw,rectangle,label=above:$i-3$,right of=h] (i) {*};
  \node[draw,rectangle,label=above:$i-4$,right of=i] (j) {*};
  \node[draw,rectangle,label=above:$i-5$,right of=j] (k) {*};
  \node[inner sep=0,minimum size=0,right of=k] (l) {$\cdots$}; 
  \node[draw,rectangle,label=above:$0$,right of=l] (m) {*};
  \node[rectangle,draw,label=left:${zs^{t}}~$,below of=a](n){0};
  \node[inner sep=0,minimum size=0,below of=b] (o) {$\cdots$}; 
  \node[rectangle,draw,below of=c](p){0};
  \node[rectangle,draw,below of=d](q){0};
  \node[rectangle,draw,below of=e](r){0};
  \node[rectangle,draw,below of=f](s){0};
  \node[rectangle,draw,below of=g](t){0};
  \node[rectangle,draw,below of=h](u){$z_{2}$};
  \node[rectangle,draw,below of=i](v){$z_{1}$};
  \node[rectangle,draw,below of=j](w){$z_{0}$};
  \node[rectangle,draw,below of=k](x){0};
  \node[inner sep=0,minimum size=0,below of=l] (y) {$\cdots$}; 
  \node[rectangle,draw,below of=m](z){0};
  \node[rectangle,draw,label=left:${u}~$,below of=n](aa){*};
  \node[inner sep=0,minimum size=0,below of=o] (bb) {$\cdots$}; 
  \node[rectangle,draw,below of=p](cc){*};
  \node[rectangle,draw,below of=q](dd){0};
  \node[rectangle,draw,below of=r](ee){0};
  \node[rectangle,draw,below of=s](ff){1};
  \node[rectangle,draw,below of=t](gg){0};
  \node[rectangle,draw,below of=u](hh){?};
  \node[rectangle,draw,below of=v](ii){?};
  \node[rectangle,draw,below of=w](jj){?};
  \node[rectangle,draw,below of=x](kk){*};
  \node[inner sep=0,minimum size=0,below of=y] (ll) {$\cdots$}; 
  \node[rectangle,draw,below of=z](mm){*};
\end{tikzpicture}
\caption{The $s$-ary expansions of $a$, $zs^{t}$, and $u = a+zs^{t}$ in \clref{case1} when $m=3$ and $t=i-4$. Observe that $a \in S_{i}$ because $a_{i+2} = a_{i+1} = a_{i} = 0$ and $u \in S_{i,t}$ because $u_{i+2} = u_{i+1}=0$, $u_{i} = 1$, and $u_{i-1} = 0$.}
\label{fig:case1}
\end{center}
\end{figure*}

\begin{figure*}[t]
\centering
\captionsetup{justification=centering}
\begin{center}
\begin{tikzpicture}[thick]
  \node[rectangle,draw,label=left:${a}~$,label=above:$n+2$](a){0};
  \node[rectangle,draw,label=above:$n+1$,right of =a](b){*};
  \node[inner sep=0,minimum size=0,right of=b] (c) {$\cdots$}; 
  \node[rectangle,draw,label=above:$i+6$,right of =c](d){*};
  \node[draw,rectangle,label=above:$i+5$,right of=d] (e) {*};
  \node[draw,rectangle,label=above:$i+4$,right of=e] (f) {*};
  \node[draw,rectangle,label=above:$i+3$,right of=f] (g) {*};
  \node[draw,rectangle,label=above:$i+2$,right of=g] (h) {0};
  \node[draw,rectangle,label=above:$i+1$,right of=h] (i) {0};
  \node[draw,rectangle,label=above:$i$,right of=i] (j) {0};
  \node[draw,rectangle,label=above:$i-1$,right of=j] (k) {*};
  \node[inner sep=0,minimum size=0,right of=k] (l) {$\cdots$}; 
  \node[draw,rectangle,label=above:$0$,right of=l] (m) {*};
  \node[rectangle,draw,label=left:${zs^{t}}~$,below of=a](n){0};
  \node[rectangle,draw,below of=b](o){0};
  \node[inner sep=0,minimum size=0,below of=c] (p) {$\cdots$}; 
  \node[rectangle,draw,below of=d](q){0};
  \node[rectangle,draw,below of=e](r){$z_{2}$};
  \node[rectangle,draw,below of=f](s){$z_{1}$};
  \node[rectangle,draw,below of=g](t){$z_{0}$};
  \node[rectangle,draw,below of=h](u){$0$};
  \node[rectangle,draw,below of=i](v){0};
  \node[rectangle,draw,below of=j](w){0};
  \node[rectangle,draw,below of=k](x){0};
  \node[inner sep=0,minimum size=0,below of=l] (y) {$\cdots$}; 
  \node[rectangle,draw,below of=m](z){0};
  \node[rectangle,draw,label=left:${u}~$,below of=n](aa){1};
  \node[rectangle,draw,below of=o](bb){0};
  \node[inner sep=0,minimum size=0,below of=p] (cc) {$\cdots$}; 
  \node[rectangle,draw,below of=q](dd){0};
  \node[rectangle,draw,below of=r](ee){?};
  \node[rectangle,draw,below of=s](ff){?};
  \node[rectangle,draw,below of=t](gg){?};
  \node[rectangle,draw,below of=u](hh){0};
  \node[rectangle,draw,below of=v](ii){0};
  \node[rectangle,draw,below of=w](jj){0};
  \node[rectangle,draw,below of=x](kk){*};
  \node[inner sep=0,minimum size=0,below of=y] (ll) {$\cdots$}; 
  \node[rectangle,draw,below of=z](mm){*};
\end{tikzpicture}
\caption{The $s$-ary expansions of $a$, $zs^{t}$, and $u = a+zs^{t}$ in \clref{case4} when $m=3$ and $t=i+3$. Note that $a \in S_{i}$ as $a_{i+2} = a_{i+1} = a_{i} = 0$ and $u \in S'_{i,t}$ as $u_{n+2} = 1$, $u_{n+1} = \cdots = u_{i+6} = u_{i+2} = u_{i+1} = u_{i} = 0$.}
\label{fig:case4}
\end{center}
\end{figure*}

\noindent where $a_{j} \in [0, s-1]$ for $j \in [0, m+n-2]$. Define the set $S_{i}$ for $i \in [0, n-1]$ by  
\begin{equation*}
S_{i} = \{a \in [0,l-1] : a_{i} = \cdots = a_{i+m-1} = 0 \},  
\end{equation*}
and define the set of $l$ polynomials 
\begin{equation*}
\{f_{i,j}\}_{j=1}^{l} = \{ \beta^{a}x^{z} : a \in S_{i}, z \in [0,n-k-1] \}.
\end{equation*}
Notice that, for each $i$, we have indeed
defined $l = s^{m+n-1}$ polynomials because
$|S_{i}| = s^{n-1}$ and because there are $n-k = s^{m}$ choices for $z$.

\begin{claim} 
\label{clm:basispoints}
For $i \in [0,n-1]$, the set $\{f_{i,1}(\beta^{s^{i}}), \ldots, f_{i,l}(\beta^{s^{i}})\}$ is a basis for $E$ over $F$. 
\end{claim}

\noindent \textbf{Proof.} 
Since $n-k = s^{m}$ and $z \in [0,n-k-1]$, we can write the $s$-ary expansion of $z$ as 
\begin{equation*}
z = \sum_{j=0}^{m-1} z_{j}s^{j},
\end{equation*}
where $z_{j} \in [0,s-1]$ for $j \in [0,m-1]$. Consequently,
\begin{equation*}
\beta^{a}(\beta^{s^{i}})^{z} = \beta^{a}(\beta^{s^{i}})^{\sum_{j=0}^{m-1} z_{j}s^{j}} = \beta^{a + \sum_{j=0}^{m-1} z_{j}s^{i+j}}.
\end{equation*}

By considering $s$-ary expansions, we see that as $a$ ranges over
$S_{i}$ and $z_{j}$ ranges over $[0,s-1]$ for \mbox{$j \in [0,m-1]$}, 
the sum $a + \sum_{j=0}^{m-1} z_{j}s^{i+j}$ ranges over $[0,l-1]$. We
thus have
\begin{equation*}
\{f_{i,1}(\beta^{s^{i}}), \ldots, f_{i,l}(\beta^{s^{i}})\} = \{1, \beta, \ldots, \beta^{l-1}\}\ .
\end{equation*}
The latter set is clearly a basis for $E$ over $F$. \qed

Our remaining task is to show that \eqref{sumofdimensions} 
is bounded from above by \eqref{weakbwbound}. 
To this end, we now establish upper bounds on
$\dim_{F}(\{f_{i,j}(\beta^{s^{t}})\}_{j=1}^{l})$ in two cases
according to whether $t \leq i-m$ or $t \geq i+m$. The case
where $t \in [i-(m-1),i+(m-1)]$ is dealt with separately.

\begin{claim}
\label{clm:case1}
If\/ $i-t \geq m$, then 
\[
\dim_{F}\left(\{f_{i,j}(\beta^{s^{t}})\}_{j=1}^{l}\right) 
\leq 
\frac{l}{n-k} + \frac{l}{s^{i-t}}.
\]
\end{claim}

\noindent \textbf{Proof.}  
Define the set
\begin{align}
S_{i,t} = \{&u \in [0,l-1] : \, u_{i+m-1} = \cdots = u_{i+1} = 0, 
u_{i} = 1, u_{i-1} = \cdots = u_{t+m} = 0\}. 
\label{carries1} 
\end{align}
Referring to Figure\,\ref{fig:case1},
we claim that if $i-t \geq m$, then 
\begin{align*}
\{f_{i,j}(\beta^{s^{t}})\}_{j=1}^{l} &= \{\beta^{a+zs^{t}} : a \in S_{i}, z \in [0,n-k-1]\} \subseteq \{\beta^{u} : u \in S_{i} \cup S_{i,t}\}.
\end{align*}
By considering the $s$-ary expansions of $a$ and $zs^{t}$, we see that
if $u = a+zs^{t} \notin S_{i}$, then the addition must have generated a carry from coordinate $t+m-1$ to coordinate $i$, which motivates \eqref{carries1}.
\clref{case1} follows because
\begin{equation*}
|S_{i}| = \frac{l}{n-k} 
\hspace{3ex}\text{and}\hspace{3ex}
|S_{i,t}| = \frac{l}{s^{i-t}}. \qed
\end{equation*}

\begin{claim}
\label{clm:case4}
If\/ $t-i \geq m$, then 
\[
\dim_{F}\left(\{f_{i,j}(\beta^{s^{t}})\}_{j=1}^{l}\right) \leq \frac{l}{n-k} + \frac{l}{s^{m+n-t-1}}.
\]
\end{claim}

\noindent 
\textbf{Proof.} 
Define the set
\begin{align}
S'_{i,t} = \{&u \in [0,s^{m+n}-1] : \, u_{m+n-1} = 1, 
u_{m+n-2} = \cdots = u_{m+t} = 0, \label{carries4}  
u_{i+m-1} = \cdots = u_{i} = 0\}. 
\end{align}
Referring to Figure\,\ref{fig:case4},
we claim that if $t-i \geq m$, then 
\begin{align*}
\{f_{i,j}(\beta^{s^{t}})\}_{j=1}^{l} &= \{\beta^{a+zs^{t}} : a \in S_{i}, z \in [0,n-k-1]\} \subseteq \{\beta^{u} : u \in S_{i} \cup S'_{i,t}\}.
\end{align*}
By considering the $s$-ary expansions of $a$ and $zs^{t}$, we see that if $u = a+zs^{t} \notin S_{i}$, then the addition must have generated a carry from coordinate $m+t-1$ to coordinate $m+n-1$, which motivates \eqref{carries4}. 
\clref{case4} now follows because
\begin{equation*}
|S'_{i,t}| = \frac{l}{s^{m+n-t-1}}. \qed
\end{equation*}
  
Finally, we bound \eqref{sumofdimensions} from above by \eqref{weakbwbound}.
An upper bound on \eqref{sumofdimensions} is
\begin{align}
\frac{\Bigl(n-2(m-1) + 2(m-1)(n-k)\Bigl)l}{n-k} 
~+~
\sum_{t=0}^{i-m} \frac{l}{s^{i-t}} 
~+
\sum_{t=i+m}^{n-1} \frac{l}{s^{m+n-t-1}}. 
\label{weaksumupcases}
\end{align}
If $i < m$ or if $i > n-1-m$, then empty sums in
\eqref{weaksumupcases} are taken as zero by convention.
The first term in \eqref{weaksumupcases} comes from summing 
the $l/(n-k)$ terms in \clref{case1} and \clref{case4} 
and using the trivial bound 
\begin{equation}
\label{weakcase2and3}
\dim_{F}\left(\{f_{i,j}(\beta^{s^{t}})\}_{j=1}^{l}\right) \leq l
\end{equation}
for all $t \in [i-(m-1),i+(m-1)]$. The second and third terms 
in \eqref{weaksumupcases} come from \clref{case1} and \clref{case4},
respectively.
Using well-known formulas for geometric series, along with the 
fact that $s \ge 2$, we can bound the second term in \eqref{weaksumupcases} 
from above by
\begin{equation}
\label{case1and2}
\sum_{t=1}^{i-m} \frac{l}{s^{i-t}} < \frac{2l}{s^m}.
\end{equation}
In a similar manner, we can bound the last term in \eqref{weaksumupcases} by
\begin{equation}
\label{case4sum}
\sum_{t=i+m}^{n-1} \frac{l}{s^{m+n-t-1}} < \frac{2l}{s^{m}}.
\end{equation}
Summing the first term of \eqref{weaksumupcases} and the
right-hand-sides of \eqref{case1and2} and \eqref{case4sum} shows that
\eqref{sumofdimensions} is bounded from above by \eqref{weakbwbound}. This
completes the proof of \tref{weakmain}. \qed

\vspace{3.0ex}
\section{Optimal-Bandwidth and Optimal-Update Codes}
\label{sec:update}

Recall that, in \cite{YB}, Ye and Barg construct optimal-bandwidth and
optimal-update $(n,k,r^{n})$ MDS codes with diagonal encoding
matrices. In this section, we adapt this construction and improve
the subpacketization at the cost of an asymptotically negligible
increase in repair bandwidth.

Table\,\ref{tab:OUtradeoff} compares our optimal-update MDS code
construction with that of Ye and Barg \cite{YB}.
If $r = 2^m$, for example, we achieve
the subpacketization of~$2^{m+n-1}$, which improves upon the
subpacketization of $2^{mn}$ in the constructions by Ye and Barg. 

In the Ye--Barg construction \cite{YB}, optimal repair bandwidth is achieved because each surviving node needs to transmit one scalar in $F$ to recover $r$ coordinates in the failed node. Our construction achieves asymptotically optimal repair bandwidth because all but a constant number of surviving nodes need to transmit one scalar in $F$ to repair $r$ coordinates in the failed node. 

In an MDS code, each parity node is a function of the entire
information stored in the system. Consequently, when any information
element changes its value, each parity node needs to update 
{at least one} of its elements.\linebreak
Here, we regard (the content stored at) each node as a vector 
in $F^l$ and refer to symbols in $F$ as its 
\textit{elements}.
An \textit{optimal-update code} is one in which
each parity node needs to update \textit{exactly one} of its elements when an
information element changes. Optimal-update codes are desirable
since updating is a frequent operation.

One way to construct optimal-update MDS codes is to encode the parity
nodes with diagonal encoding~matrices. In other words, each parity
node $C_{k+i} \in F^{l}$ for $i \in [r]$ is defined by
\begin{equation}
\label{parityencoding}
C_{k+i} = \sum_{j=1}^{k} D_{i,j}C_{j},
\end{equation}
where $C_{1}, \ldots, C_{k} \in F^{l}$ are the systematic nodes and
$D_{i,j}$ is an $l \times l$ diagonal matrix. In \cite{YB}, Ye and
Barg~construct optimal-bandwidth and optimal-update $(n,k,r^{n})$ MDS
codes with diagonal encoding matrices.\pagebreak[3.99]

\looseness=-1
We adapt the Ye--Barg construction in \cnref{adapt} and improve the
subpacketization at the expense~of~an asymptotically negligible increase
in the repair bandwidth. \tref{bb} shows that the resulting code has
asymptotically optimal repair bandwidth while being optimal-update.
When $r$ is not an integral power, we
show how to obtain optimal update MDS codes with near-optimal
bandwidth in \cref{bbnopower}.

Note that, with our level of subpacketization, our MDS codes cannot meet the cut-set bound \eqref{cutset} because~a~result of \cite{TWB} shows that an optimal-bandwidth $(n,k,l)$ MDS code with diagonal encoding matrices satisfies $l \geq r^{k}$. For fixed $r=s^{m}$, as $n \rightarrow \infty$, we have $s^{m+n-1} < r^{k}$, so the most we can hope for is asymptotically optimal repair bandwidth.

Let $\cC$ be an $(n,k,l)$ MDS code with nodes $C_{i} \in F^{l}$ represented as column vectors for $i \in [n]$. We consider codes defined in the following parity-check form
\begin{equation}
\label{paritycheckform}
\cC = \left\{(C_{1}, \ldots, C_{n}) : \sum_{i=1}^{n} A_{t,i}C_{i} = \zero, t \in [r] \right\},
\end{equation}
where $A_{t,i}$ is an $l \times l$ matrix over $F$ for $t \in [r]$ and $i \in [n]$. Given positive integers $r$ and $n$, define an $(n, k, l)$ MDS code $\cC$ by setting in \eqref{paritycheckform}
\begin{equation}
\label{matrixpowers}
A_{t,i} = A_{i}^{t-1}, t \in [r], i \in [n],
\end{equation}
where $A_{1}, \ldots, A_{n}$ are $l \times l$ matrices that will be specified in \cnref{adapt}. We use the convention $A^{0} = I$. 

\begin{definition}
\label{def:sarytorary}
Let $s \geq 2$ and $m \geq 1$ be positive integers. Let $l = s^{m+n-1}$. Given $a \in [0,l-1]$, we can write its $s$-ary expansion as $(a_{m+n-1}, \ldots, a_{1})$; that is
\[
a = \sum_{j=1}^{m+n-1} a_{j}s^{j-1},
\]
where $a_{j} \in [0,s-1]$ for $j \in [m+n-1]$. For $i \in [n]$, let $a_{(i+m-1, \ldots, i)}$ be the unique $x \in [0,s^{m}-1]$ such that the $s$-ary representation of $x$ is $(a_{i+m-1}, \ldots, a_{i})$.
\end{definition}

To illustrate \dref{sarytorary} consider the following example. 

\vspace{0.1cm}

\begin{example} 
\label{ex:exsarytorary}
Let $s=2$, $m=2$, $n=10$, and $l = 2^{11}$. Let $a = 6$ whose binary expansion is $(0, \ldots, 0,1,1,0)$. We have
\[
6_{(2,1)} = 2, \; 6_{(3,2)} = 3, \; 6_{(4,3)} = 1, \; 6_{(i+1,i)} = 0 \; \mbox{if $i \in [4,10]$}.   
\]
\end{example}

We now show how to adapt Construction 1 of Ye and Barg in \cite{YB}. 

\begin{construction}
\label{cnstr:adapt}
Let $n$ and $k$ be fixed integers and suppose that $n-k \geq s^{m}$, where $s \geq 2$ and $m \geq 1$. Let $F$ be a finite field of size $|F| \geq s^{m}n$ and let $l = s^{m+n-1}$. Let $\{ \lambda_{i,j} \}_{i \in [n], j \in [0,s^{m}-1]}$ be $s^{m}n$ distinct elements in $F$. Consider the code family given by \eqref{paritycheckform} and \eqref{matrixpowers} where we take
\[
A_{i} = \sum_{a=0}^{l-1} \lambda_{i,a_{(i+m-1, \ldots, i)}} e_{a}e_{a}^{\top}, \; \; i \in [n].
\]
Here, $\{e_0,\ldots,e_{l-1}\}$ is the standard basis 
for $F^{l}$ over $F$, viewed as column vectors.
\end{construction}

Since the $A_{i}$ for $i \in [n]$ are diagonal matrices, we can write out the parity-check equations coordinatewise. Letting $c_{i,a}$ denote the $a^{\mathrm{th}}$ coordinate of the column vector $C_{i}$, we have for all integers $a \in [0, l-1]$ and $t \in [0,r-1]$ that
\begin{equation}
\label{coordinatewise}
\sum_{i=1}^{n} \lambda^{t}_{i, a_{(i+m-1, \ldots, i)}}c_{i,a} = 0.
\end{equation}

To illustrate \eqref{coordinatewise} consider the following example. 

\vspace{0.1cm}

\begin{example}
\label{ex:excoordinatewise}
Suppose $s=2$, $m=2$, $r=4$, $n=10$, and $l = 2^{11}$. For $a \in \{0,2,4,6\}$ and $t \in [0,3]$, the equations in \eqref{coordinatewise} are
\begin{align*}
\lambda^{t}_{1,0}c_{1,0} + \lambda^{t}_{2,0}c_{2,0} + \lambda^{t}_{3,0}c_{3,0} + \lambda^{t}_{4,0}c_{4,0} + \sum_{j=5}^{10} \lambda^{t}_{j,0}c_{j,0} &= 0. \notag \\
\lambda^{t}_{1,2}c_{1,2} + \lambda^{t}_{2,1}c_{2,2} + \lambda^{t}_{3,0}c_{3,2} + \lambda^{t}_{4,0}c_{4,2} + \sum_{j=5}^{10} \lambda^{t}_{j,0}c_{j,2} &= 0. \notag \\
\lambda^{t}_{1,0}c_{1,4} + \lambda^{t}_{2,2}c_{2,4} + \lambda^{t}_{3,1}c_{3,4} + \lambda^{t}_{4,0}c_{4,4} + \sum_{j=5}^{10} \lambda^{t}_{j,4}c_{j,4} &= 0. \notag \\
\lambda^{t}_{1,2}c_{1,6} + \lambda^{t}_{2,3}c_{2,6} + \lambda^{t}_{3,1}c_{3,6} + \lambda^{t}_{4,0}c_{4,6} + \sum_{j=5}^{10} \lambda^{t}_{j,6}c_{j,6} &= 0. \notag \\
\end{align*}
\end{example}

\tref{bb} addresses the repair bandwidth of the code in \cnref{adapt} when $r=s^{m}$. The full proof of \tref{bb} is presented in \aref{appendixb}. 

\begin{theorem}
\label{thm:bb}
If $n-k=s^{m}$, the exact repair bandwidth of the code in \cnref{adapt} is at most
\begin{equation}
\label{bandwidthbound}
\left( \frac{n-1 + 2 \sum_{v=1}^{m-1} (s^{v}-1)}{n-k} \right)l,
\end{equation}
and hence asymptotically meets \eqref{cutset} for fixed $n-k$ as $n \to \infty$.
\end{theorem}

As in \cref{nopower}, even if $r$ is not an integral power, we can still use the ideas in \tref{bb}. We can choose any positive integers $s$ and $m$ such that $s^{m} \leq r$. The statement and proof of \tref{bb} still hold if we replace by $s^{m}$ every occurrence of $n-k$. For fixed $r$, as $n \rightarrow \infty$, the ratio between the repair bandwidth of the resulting codes and \eqref{cutset} would be $r/s^{m}$, which is at most $2$. We summarize this result in the following corollary.

\begin{corollary}
\label{cor:bbnopower}
If $n-k \geq s^{m}$, the exact repair bandwidth of the code in \cnref{adapt} is at most
\begin{equation}
\label{bandwidthboundnopower}
\left( \frac{n-1 + 2 \sum_{v=1}^{m-1} (s^{v}-1)}{s^{m}} \right)l.
\end{equation}
For fixed $n-k$, as $n \rightarrow \infty$, the ratio between \eqref{bandwidthboundnopower} and \eqref{cutset} would be $(n-k)/s^{m}$, which is at most $2$.
\end{corollary}

In this section, we prove \tref{weakbb}, a version of \tref{bb} with a slightly weaker bound on the repair bandwidth. In conjunction with the proof of \tref{weakbb}, the reader may find it useful to consult \eref{exweakbb}, which illustrates the notation in \tref{weakbb}.

\begin{theorem}
\label{thm:weakbb}
If $n-k = s^{m}$, the exact repair bandwidth of the code in \cnref{adapt} is at most
\begin{equation}
\label{weakbandwidthbound}
\left( \frac{n-1 + 2(m-1)s^{m} - 2m + 2}{n-k} \right)l,
\end{equation}
and hence asymptotically meets \eqref{cutset} for fixed $n-k$ as $n \to \infty$.
\end{theorem}

\noindent \textbf{Proof.} For $(w_{m}, \ldots, w_{1}) \in [0,s-1]^{m}$ and $a \in [0,l-1]$, define $a(i; w_{m}, \ldots, w_{1}) \in [0,l-1]$ to be the integer whose $s$-ary representation is obtained from the $s$-ary representation of $a$ by replacing the $m$ coordinates $a_{i+m-1}, \ldots, a_{i}$ with $w_{m}, \ldots, w_{1}$ respectively; that is
\begin{align}
&a(i; w_{m}, \ldots, w_{1}) \notag \\
&= (a_{n+m-1}, \ldots, a_{i+m}, w_{m}, \ldots, w_{1}, a_{i-1}, \ldots, a_{1}). \label{varying}
\end{align}
For $i \in [n]$ and $a \in [0,l-1]$, let $S_{a,i}$ be the set of integers in $[0,l-1]$ whose $s$-ary representation is obtained from the $s$-ary representation of $a$ by replacing the $m$ coordinates $a_{i+m-1}, \ldots, a_{i}$ with $w_{m}, \ldots, w_{1}$ respectively for all $(w_{m}, \ldots, w_{1}) \in [0,s-1]^{m}$; that is
\begin{equation}
\label{Sidef}
S_{a,i} = \{a(i; w_{m}, \ldots, w_{1}) : (w_{m}, \ldots, w_{1}) \in [0,s-1]^{m} \}.
\end{equation}
For $a \in [0,l-1]$ and $j \in [n] \setminus \{i\}$, we define elements in $F$ by 
\begin{equation}
\label{ujia} 
u^{(a)}_{j,i} = \sum_{a' \in S_{a,i}} c_{j,a'}.
\end{equation}
  
We will show that for any $i \in [n]$ and $a \in [0,l-1]$, the $r$ coordinates
\begin{equation}
\label{rcoords}
\{c_{i, a'} : a' \in S_{a,i}\}
\end{equation}
in $C_{i}$ are functions of the following set $D^{(a)}_{i} \subset F$,
\begin{equation*}
D^{(a)}_{i} = \{u^{(a)}_{j,i} : |j-i| \geq m\} \sqcup \{c_{j,a'}: |j-i| < m, a' \in S_{a,i}\}.
\end{equation*}
To repair $r$ coordinates in the failed node, a surviving node $j$ needs to transmit one scalar in $F$ if $|j-i| \geq m$ and $r$ scalars otherwise. Consequently, 
\begin{equation*}
|D^{(a)}_{i}| \leq (n-1)-2(m-1)+2(m-1)r,
\end{equation*}
and so the repair bandwidth of the code in \cnref{adapt} is at most \eqref{weakbandwidthbound}.

We write \eqref{coordinatewise} for $t \in [0,r-1]$ and sum over $a' \in S_{a,i}$. When $t=0$, we obtain
\begin{equation}
\label{tis0}
\sum_{a' \in S_{a,i}} c_{i, a'} = -\sum_{j \neq i} u^{(a)}_{j,i}.
\end{equation}
Notice that if $|j-i| \geq m$, then for all $a' \in S_{a,i}$, the value of $a'(j+m-1, \ldots, j)$ is the same. For $j$ such that $|j-i| \geq m$ define $l_{a,j} \in [0,r-1]$ to be the value of $a'(j+m-1, \ldots, j)$ for any $a' \in S_{a,i}$. When $1 \leq t \leq r-1$, we obtain
\begin{align}
\label{tisnotzeroweak}
&\sum_{a' \in S_{a,i}} \lambda^{t}_{i, a'_{(i+m-1, \ldots, i)}} c_{i, a'} = - \sum_{j : |j-i| \geq m} \lambda^{t}_{j, l_{a,j}} u^{(a)}_{j,i} \notag \\
&- \sum_{j : |j-i| < m} \sum_{a' \in S_{a,i}} \lambda^{t}_{j, a'_{(j+m-1, \ldots, j)}} c_{j,a'}.
\end{align}
For $t=0$, let $B_{0}$ denote the right-hand-side of \eqref{tis0}. Similarly, for $1 \leq t \leq r-1$, let $B_{t}$ denote the right-hand-side of \eqref{tisnotzeroweak}. Writing the system of equations given by \eqref{tis0} and \eqref{tisnotzeroweak} in matrix form yields
\begin{equation}
\label{matrixform}
\left[\begin{array}{cccc}
       1 & \cdots & 1 \\
       \lambda_{i,0} & \cdots & \lambda_{i,r-1} \\
       \vdots & \vdots & \vdots \\
       \lambda_{i,0}^{r-1} & \cdots & \lambda_{i,r-1}^{r-1}
      \end{array}\right]
\left[\begin{array}{c}
       c_{i,a(i; 0, \ldots, 0)} \\
       \vdots \\
       c_{i,a(i; s-1, \ldots, s-1)}
      \end{array}\right]
= \left[\begin{array}{c}
         B_{0} \\
         \vdots \\
         B_{r-1} 
        \end{array}\right].      
\end{equation}
Since $B_{0}, \ldots, B_{r-1}$ are functions of the elements in $D^{(a)}_{i}$ and since $\lambda_{i,0}, \ldots, \lambda_{i,r-1}$ are distinct, we can solve the system in \eqref{matrixform} for the $r$ coordinates in \eqref{rcoords} given $D^{(a)}_{i}$. \qed

To illustrate \tref{weakbb}, consider the following example.

\vspace{0.1cm}

\begin{example}
\label{ex:exweakbb}
Suppose $s=2$, $m=2$, $r=4$, $n=10$, and $l=2^{11}$. Suppose node 2 has failed so $i=2$. Letting $a=0$, we have in \eqref{varying} that
\[
0(2;0,0) = 0, \, 0(2;0,1) = 2, \, 0(2;1,0) = 4, \, 0(2;1,1) = 6.
\]
Hence, $S_{0,2} = \{0,2,4,6\}$ and 
\begin{equation}
\label{ujiaex}
u^{(0)}_{j,2} = c_{j,0} + c_{j,2} + c_{j,4} + c_{j,6}.
\end{equation}
The four coordinates in \eqref{rcoords} are $\{c_{2,0}, c_{2,2}, c_{2,4}, c_{2,6}\}$ and we claim these coordinates are functions of the set
\begin{align*}
&D^{(0)}_{2} = \{ u^{(0)}_{j,2} : 4 \leq j \leq 10 \} \; \cup \notag \\
&\{c_{j,a'}: j \in \{1,3\}, a' \in \{0,2,4,6\} \}. 
\end{align*}
To see why the coordinates $c_{2,0}, c_{2,2}, c_{2,4}, c_{2,6}$ are functions of the set 
$D^{(0)}_{2}$, sum the equations in \eref{excoordinatewise}. When $t = 0$, we obtain
\begin{equation}
\label{tis0ex}
c_{2,0} + c_{2,2} + c_{2,4} + c_{2,6} = -\sum_{j \neq 2} u^{(0)}_{j,2}.
\end{equation}
We have $l_{0,j} = 0$ for $4 \leq j \leq 10$, so when $1 \leq t \leq 3$, we obtain
\begin{align*}
&\lambda^{t}_{2,0}c_{2,0} + \lambda^{t}_{2,1}c_{2,2} + \lambda^{t}_{2,2}c_{2,4} + \lambda^{t}_{2,3}c_{2,6} \notag \\
&= -\lambda^{t}_{1,0}(c_{1,0} + c_{1,4}) - \lambda^{t}_{1,2}(c_{1,2} + c_{1,6}) - \lambda^{t}_{3,0}(c_{3,0} + c_{3,2}) \notag \\
&- \lambda^{t}_{3,1}(c_{3,4} + c_{3,6}) -\sum_{j=4}^{10} \lambda^{t}_{j,0} u^{(0)}_{j,2}. 
\end{align*}
\end{example}

We now show that the code in \cnref{adapt} is MDS.

\begin{theorem}
\label{thm:codeisMDS}
The code $\cC$ given by \cnref{adapt} is MDS.
\end{theorem}

\noindent \textbf{Proof.} Writing the parity check equations \eqref{paritycheckform} coordinatewise, we have for all $a \in [0,l-1]$ that
\begin{equation}
\label{paritycheckcw}
\left[\begin{array}{cccc}
       1 & \cdots & 1 \\
       \lambda_{1,a(m,\ldots,1)} & \cdots & \lambda_{n,a(n+m-1, \ldots, n)} \\
       \vdots & \vdots & \vdots \\
       \lambda^{r-1}_{1,a(m,\ldots,1)} & \cdots & \lambda^{r-1}_{n,a(n+m-1, \ldots, n)}
      \end{array}\right]
\left[\begin{array}{c}
       c_{1,a} \\
       \vdots \\
       c_{n,a}
      \end{array}\right]
= \zero.
\end{equation}
Since every $r$ columns of the parity-check matrix in \eqref{paritycheckcw} have rank $r$, any $k$ out of $n$ elements in the set $\{c_{1,a} \ldots, c_{n,a}\}$ can recover the whole set. As this holds for all $a \in [0,l-1]$, any $k$ nodes of a codeword in $\cC$ can recover the whole codeword. \qed

Finally, we show that the code constructed in \cnref{adapt} has diagonal encoding matrices and is thus optimal-update. 

\begin{theorem}
\label{thm:diagonal}
The code $\cC$ given by \cnref{adapt} has diagonal encoding matrices and is thus optimal-update.
\end{theorem}

\noindent \textbf{Proof.} We assume that the first $k$ nodes $C_{1}, \ldots, C_{k}$ are the systematic nodes, and we seek diagonal encoding matrices $D_{i,j}$ satisfying \eqref{parityencoding}. Let
\begin{align*}
V_{1}^{(a)} &= \left[\begin{array}{cccc}
               1 & \cdots & 1 \\
               \lambda_{k+1,a(k+m,\ldots,k+1)} & \cdots & \lambda_{n,a(n+m-1, \ldots, n)} \\
               \vdots & \vdots & \vdots \\
               \lambda^{r-1}_{k+1,a(k+m,\ldots,k+1)} & \cdots & \lambda^{r-1}_{n,a(n+m-1, \ldots, n)}
               \end{array}\right] 
               \\ 
V_{2}^{(a)} &= \left[\begin{array}{cccc}
              1 & \cdots & 1 \\
              \lambda_{1,a(m,\ldots,1)} & \cdots & \lambda_{k,a(k+m-1, \ldots, k)} \\
              \vdots & \vdots & \vdots \\
              \lambda^{r-1}_{1,a(m,\ldots,1)} & \cdots & \lambda^{r-1}_{k,a(k+m-1, \ldots, k)}
              \end{array}\right]. 
\end{align*}
By \eqref{paritycheckcw}, we have 
\begin{equation*}
V_{1}^{(a)} \left[\begin{array}{c}
       c_{k+1,a} \\
       \vdots \\
       c_{n,a}
      \end{array}\right]
= -V_{2}^{(a)} \left[\begin{array}{c}
       c_{1,a} \\
       \vdots \\
       c_{k,a}
      \end{array}\right],
\end{equation*}
or equivalently,
\begin{equation*}
\left[\begin{array}{c}
       c_{k+1,a} \\
       \vdots \\
       c_{n,a}
      \end{array}\right]
=-\left(V_{1}^{(a)}\right)^{-1}V_{2}^{(a)}\left[\begin{array}{c}
       c_{1,a} \\
       \vdots \\
       c_{k,a}
      \end{array}\right].
\end{equation*}
Letting $M^{(a)}=-\left(V_{1}^{(a)}\right)^{-1}V_{2}^{(a)}$, we have for each $i \in [r]$ and $a \in [0,l-1]$ that
\begin{equation*}
c_{k+i,a} = M^{(a)}(i,1)c_{1,a} + \cdots + M^{(a)}(i,k)c_{k,a}, 
\end{equation*}
where $M^{(a)}(i,j)$ is the $(i,j)$ entry of the matrix $M^{(a)}$. Defining an $l \times l$ diagonal matrix $D_{i,j}$ by
\begin{equation*}
D_{i,j}(a,a) = M^{(a)}(i,j),
\end{equation*}
we have that \eqref{parityencoding} is satisfied. \qed


\section{A Framework for Constructing Optimal-Bandwidth MDS Codes}
\label{sec:framework}

In \cite{TWB}, Tamo, Wang, and Bruck show that constructing optimal-bandwidth $(n,k,l)$ MDS codes is equivalent to finding encoding matrices and repairing subspaces that satisfy certain properties. In this section, we adapt their framework for use in constructing $(n,k,l)$ MDS codes that have repair-by-transfer schemes with asymptotically optimal repair bandwidth. 

We represent the $n$ nodes $C_{1}, \ldots, C_{n} \in F^{l}$ as column vectors and for $a \in [0,l-1]$, we let $c_{i,a}$ denote the $a^{\mathrm{th}}$ coordinate of $C_{i}$. We assume that the first $k$ nodes $C_{1}, \ldots, C_{k}$ are systematic and that the parity nodes $C_{k+i}$ for $i \in [r]$ are defined by
\begin{equation}
\label{parityencoder}
C_{k+i} = \sum_{j=1}^{k} A_{i,j}C_{j},
\end{equation}
where $A_{i,j}$ is an invertible $l \times l$ \textit{encoding matrix}. Hence, the code is uniquely determined by the matrix
\begin{equation}
\label{codematrix}
\A = (A_{i,j})_{i \in [r], j \in [k]} = \left[\begin{array}{ccc}
                                                A_{1,1} & \cdots & A_{1,k} \\
                                                \vdots & \vdots & \vdots \\
                                                A_{r,1} & \cdots & A_{r,k}
                                               \end{array}\right].
\end{equation}
The code is MDS if and only if any $1 \times 1, 2 \times 2, \ldots, r \times r$ block submatrix of $\A$ is invertible. 

We will make some additional assumptions on the encoding matrices $A_{i,j}$. Denote the matrix $A_{2,j}$ by $A_{j}$. We will assume the matrix $A_{j}$ is diagonalizable and has $r$ distinct nonzero eigenvalues corresponding to $r$ eigenspaces, each of dimension $l/r$. We define the matrix $A_{i,j} = A_{j}^{i-1}$ for $i \in [r]$ and $j \in [k]$. Consequently, all the matrices $A_{i,j}$ are diagonalizable. Moreover, if $V_{j,0}, \ldots, V_{j,r-1}$ are the left eigenspaces of the matrix $A_{j}$ corresponding to eigenvalues $\lambda_{j,0}, \ldots, \lambda_{j,r-1}$, then the matrix $A_{i,j}$ also has left eigenspaces $V_{j,0}, \ldots, V_{j,r-1}$ corresponding to eigenvalues $\lambda_{j,0}^{i-1}, \ldots, \lambda_{j,r-1}^{i-1}$.

For each systematic node $j \in [k]$, we will define a repairing subspace $S_{j}$ of dimension $\dim(S_{j}) = l/r$. During the repair process of systematic node $h$, parity node $k+i$ projects its data onto the repairing subspace $S_{h}$. By \eqref{parityencoder}, parity node $k+i$ for $i \in [r]$ transmits
\begin{equation}
\label{projection}
S_{h}C_{k+i} = S_{h} \left( \sum_{j=1}^{k} A_{i,j}C_{j} \right).
\end{equation}
Rearranging \eqref{projection} and recalling that $A_{i,h} = A_{h}^{i-1}$, we obtain
\begin{equation}
\label{projectionrewrite}
S_{h}A_{h}^{i-1}C_{h} = S_{h}C_{k+i} - \sum_{\substack{j = 1 \\ j \neq h}}^{k} S_{h}A_{j}^{i-1}C_{j}.
\end{equation}
Writing \eqref{projectionrewrite} for each $i \in [r]$ yields
\begin{equation}
\label{projectionsystem}
\left[\begin{array}{c}
       S_{h} \\
       S_{h}A_{h} \\
       \vdots \\
       S_{h}A_{h}^{r-1}
      \end{array}\right]C_{h} 
=
\left[\begin{array}{c}
       S_{h}C_{k+1} \\
       S_{h}C_{k+2} \\
       \vdots \\
       S_{h}C_{k+r}
      \end{array}\right]
-
\sum_{\substack{j=1 \\ j \neq h}}^{k} 
\left[\begin{array}{c}
       S_{h} \\
       S_{h}A_{j} \\
       \vdots \\
       S_{h}A_{j}^{r-1}
      \end{array}\right]C_{j}. 
\end{equation}

To compute the erased node $C_{h}$ from \eqref{projectionsystem}, the matrix on the left-hand-side of \eqref{projectionsystem} must be invertible; an equivalent condition is that
\begin{equation}
\label{rankconditionrestate}
S_{h} + S_{h}A_{h} + \cdots + S_{h}A_{h}^{r-1} = F^{l}.
\end{equation}

\begin{table*}[ht]
  \centering
  \captionsetup{justification=centering}
  \begin{tabular}{|c|c|c|c|c|c|c|c|c|}
    \hline
    Node index $i$ & 1 & 2 & 3 & 4 & 5 & 6 & 7 & 8 \\
    \hline
    Basis for $1^{\mathrm{st}}$             & $e_2$ & $e_1$ & $e_0$ & $e_0$ & $e_0$ &       $e_0$ & $e_0$ & $e_0$ \\
    eigenspace of $A_i$       & $e_3$ & $e_5$ & $e_1$ & $e_4$ & $e_1$ & $e_4$ & $e_1$ &  $e_4$ \\
    \hline
    Basis for $2^{\mathrm{nd}}$             & $e_4$ & $e_2$ & $e_4$ & $e_2$ & $e_2$ & $e_1$ & $e_2$ & $e_1$ \\
    eigenspace  of $A_i$      & $e_5$ & $e_6$ & $e_5$ & $e_6$ & $e_3$ & $e_5$ & $e_3$ & $e_5$ \\
    \hline
    Basis for $3^{\mathrm{rd}}$             & $e_6$ & $e_3$ & $e_6$ & $e_3$ & $e_6$ & $e_3$ & $e_4$ & $e_2$ \\
    eigenspace of $A_i$       & $e_7$ & $e_7$ & $e_7$ & $e_7$ & $e_7$ & $e_7$ & $e_5$ & $e_6$ \\
    \hline
    Basis for $4^{\mathrm{th}}$             & $w$ & $y$  & $w$ & $y$ & $w$ & $y$ & $w$ & $y$ \\
    eigenspace of $A_i$       & $x$ & $z$ & $x$ & $z$ & $x$ & $z$ & $x$ & $z$  \\
    \hline
   Basis for repairing        & $e_0$ & $e_0$ & $e_2$ & $e_1$ & $e_4$ & $e_2$ & $e_6$ & $e_3$ \\
    subspace $S_i$            & $e_1$ & $e_4$ & $e_3$ & $e_5$ & $e_5$ & $e_6$ & $e_7$ & $e_7$ \\
    \hline
  $S_{i}$ not invariant       &$A_{2}$ & $A_{2}$ & $A_{6}$ & $A_{1}$ & $A_{2}$ & $A_{3}$ & $A_{6}$ & $A_{3}$ \\
  subspace of                 &$A_{4}$ & $A_{5}$ & $A_{8}$ & $A_{5}$ & $A_{4}$ & $A_{7}$ & $A_{8}$ & $A_{7}$ \\
  \hline
  \end{tabular}
  \caption{An $(n=12,k=8,l=8)$ MDS code.}\label{tab:wtbadaptex}
  Define $w = e_{0} + e_{2} + e_{4} + e_{6}$, $x = e_{1} + e_{3} + e_{5} + e_{7}$, $y = e_{0} + e_{1} + e_{2} + e_{3}$, and $z = e_{4} + e_{5} + e_{6} + e_{7}$.
\end{table*}

\noindent Moreover, systematic node $j \neq h$ must transmit enough information to compute
\begin{equation}
\label{systematictransmission}
\left[\begin{array}{c}
       S_{h} \\
       S_{h}A_{j} \\
       \vdots \\
       S_{h}A_{j}^{r-1}
      \end{array}\right]C_{j}. 
\end{equation}
The code is optimal-bandwidth when systematic node $j \neq h$ transmits $l/r$ symbols or, equivalently, when the matrix on the left-hand-side of \eqref{systematictransmission} has dimension $l/r$. 
Since $\dim(S_{h}) = l/r$, the code is optimal-bandwidth when
\begin{equation}
\label{rankconditiontworewrite}
S_{h} = S_{h}A_{j} = \cdots = S_{h}A_{j}^{r-1},
\end{equation}
or equivalently when $S_{h}$ is an invariant subspace of $A_{j}$ for all $j \in [k] \setminus \{h\}$. Since $A_{j}$ is diagonalizable, $S_{h}$ is an invariant subspace of $A_{j}$ if and only if there exists a basis of $S_{h}$ consisting of eigenvectors of $A_{j}$.


Wang, Tamo, and Bruck \cite{TWB, WTBExplicitPublished} construct an optimal-bandwidth $(n,k,l)$ code by designing encoding matrices $A_{j}$ and repairing subspaces $S_{j}$ for $j \in [k]$ such that \eqref{rankconditionrestate} and \eqref{rankconditiontworewrite} hold. A similar argument yields a sufficient condition to construct $(n,k,l)$ codes with asymptotically optimal repair bandwidth.

\begin{theorem}
\label{thm:sufficientforAOB}
To construct an $(n,k,l)$ code which can repair systematic nodes with asymptotically optimal repair bandwidth, it suffices to
\begin{enumerate}
\item define diagonalizable matrices $A_{j}$ for $j \in [k]$ with $r$ distinct nonzero eigenvalues corresponding to $r$ eigenspaces, each of dimension $l/r$,
\item define repairing subspaces $S_{j}$ for $j \in [k]$ of dimension 
\[
\dim(S_{j}) = l/r,
\]
\item show that, for each $h \in [k]$, the repairing subspace $S_{h}$ is an invariant subspace of $A_{j}$ (equivalently that \eqref{rankconditiontworewrite} holds), for all but a constant number of nodes $j \in [k] \setminus \{h\}$,
\item show that, for each $h \in [k]$, \eqref{rankconditionrestate} holds.
\end{enumerate}
\end{theorem}

Moreover, we claim that an $(n,k,l)$ code satisfying \tref{sufficientforAOB} has a repair-by-transfer scheme with asymptotically optimal repair bandwidth if, for each systematic node $h \in [k]$, there exists a subset $V_{h} \subset [0,l-1]$ such that the row space of the matrix $S_{h}$ equals $\langle e_{a} : a \in V_{h} \rangle$. By the third condition in \tref{sufficientforAOB}, for each systematic node $h \in [k]$, there exists a set $K_{h} \subset [k]$ such that for all systematic nodes $j \in [k] \setminus K_{h}$, the repairing subspace $S_{h}$ is an invariant subspace of the encoding matrix $A_{j}$. If the row space of $S_{h}$ equals $\langle e_{a} : a \in V_{h} \rangle$, then to repair systematic node $h$, it suffices for a node $j \in [n] \setminus K_{h}$ to send the symbols $\{c_{j,a}: a \in V_{h} \}$. Since the size of $K_{h}$ is a constant depending only on $r$, such a repair scheme would have asymptotically optimal bandwidth because, in the worst case, all nodes in $K_{h}$ could send all their symbols. 


\section{Repair-by-Transfer Schemes with Improved Subpacketization}

\label{sec:ourcode}

When $r=s^{m}$ is an integral power, we can adapt the code construction in \cite{WTBExplicitPublished} to obtain, for any $t \in \mathbb{Z}^{+}$, a
\begin{equation}
\label{parameters}
(n = rt+r, k=rt, l = s^{m+t-1})
\end{equation}
MDS code with a repair-by-transfer scheme that can repair systematic nodes with asymptotically optimal repair bandwidth. If $r = 2^m$, for example, we achieve the subpacketization of $2^{k/r+m-1}$, which improves upon the subpacketization of $2^{mn/(r+1)}$ in the Wang--Tamo--Bruck construction. Table\,\ref{tab:tradeoff} compares our code construction with other state-of-the-art $(n,k,l)$ MDS code constructions. When $m=1$, our codes are equivalent to those in \cite{CHLM}.

\begin{definition}
\label{def:sarytoraryreverse}
Suppose that $r = s^{m}$, where $s \geq 2$ and $m \geq 1$. Let $l = s^{m+t-1}$. Given $a \in [0,l-1]$, we can write its $s$-ary expansion as $(a_{1}, \ldots, a_{m+t-1})$; that is
\[
a = \sum_{j=1}^{m+t-1} a_{j}s^{m+t-1-j}.
\]
For $i \in [t]$, let $a_{(i, \ldots, i+m-1)}$ be the unique $x \in [0,r-1]$ such that the $s$-ary representation of $x$ is $(a_{i}, \ldots, a_{i+m-1})$.
\end{definition}

To illustrate \dref{sarytoraryreverse}, consider the following example.

\vspace{0.1cm}

\begin{example}
\label{ex:exsarytoraryreverse} 
Suppose $s=2$, $m=2$, $r=4$, $t=5$, and $l=64$. Let $a=6$ whose binary expansion is $(0,0,0,1,1,0)$. We have
\[
6_{(i,i+1)} = 0 \; \mbox{for $i \in [1,2]$}, \; 6_{(3,4)} = 1, \; 6_{(4,5)} = 3, \; 6_{(5,6)} = 2. 
\]
\end{example}

For $a \in [0,l-1]$ and $i \in [t]$, we define $M_{a,i}$ to be the set of $r$ indices in $[0,l-1]$ that differ from $a$ in at most their $i^{\mathrm{th}}$, $(i+1)^{\mathrm{st}}, \ldots, (i+m-1)^{\mathrm{th}}$ digits. In other words,
\begin{align}
M_{a,i} = \{ &(x_{1}, \ldots, x_{m+t-1}) \in [0,l-1] : x_{j} = a_{j} \notag \\ &\mbox{for $j \notin [i, i+m-1]$}\}. \label{Mai} 
\end{align}
To illustrate \eqref{Mai}, consider the following example.

\vspace{0.1cm}

\begin{example}
\label{ex:exMai} 
Suppose $s=2$, $m=2$, $r=4$, $t=2$, and $l=8$. We have $M_{(0,0,0),1} = \{0,2,4,6\}$. Notice that $M_{(0,0,0),1}$ is the same set as $M_{(0,1,0),1}$, $M_{(1,0,0),1}$, and $M_{(1,1,0),1}$.
\end{example}

\vspace{0.1cm}

We now define some subspaces that will be used to define the left eigenspaces of our encoding matrices and our repairing subspaces.

\begin{definition}
\label{def:Piu}
Suppose that $r = s^{m}$, where $s \geq 2$ and $m \geq 1$. Let $l = s^{m+t-1}$. Let $\{e_{a} : a \in [0,l-1] \}$ be the standard basis of $F^{l}$ over $F$. For $i \in [t]$ and $u \in [0,r-1]$,
\begin{align*}
P_{i,u} &= \mathsf{span} \left( e_{a} : a_{(i,i+1, \ldots, i+t-1)}=u \right), \\
P_{i,r} &= \mathsf{span} \left( \sum_{a' \in M_{a,i}} e_{a'} : a \in [0, s^{m+t-1}-1] \right).
\end{align*}
\end{definition}
To illustrate \dref{Piu}, consider the following example.

\vspace{0.1cm}

\begin{example}
\label{ex:exPiu}
Suppose $s=2$, $m=2$, $r=4$, $t=2$, and $l=8$. 
\begin{align*}
P_{1,0} &= \mathsf{span} \left( e_{a} : a_{(1,2)} = 0 \right) = \langle e_{0}, \, e_{1} \rangle \\
P_{1,4} &= \langle e_{0} + e_{2} + e_{4} + e_{6}, \, e_{1} + e_{3} + e_{5} + e_{7} \rangle
\end{align*}
\end{example}

When $r=s^{m}$ is an integral power, we can adapt the code construction in \cite{WTBExplicitPublished} to obtain, for any $t \in \mathbb{Z}^{+}$, an $(n,k,l)$ MDS code with parameters given by \eqref{parameters} that has a repair-by-transfer scheme which can repair systematic nodes with asymptotically optimal repair bandwidth.

\begin{construction}
\label{cnstr:wtbadapt}
Suppose that $r = s^{m}$, where $s \geq 2$ and $m \geq 1$. Let $l = s^{m+t-1}$. Let $u \in [0,r-1]$ and let $i \in [t]$. For each integer $ut+i \in [k]$, define the encoding matrix $A_{ut+i}$. The left eigenspaces of $A_{ut+i}$ are $P_{i,u'}$ for $u' \in [0,r] \setminus \{u\}$ and correspond to distinct nonzero eigenvalues. Furthermore, let $P_{i,u}$ be the repairing subspace for node $ut+i$, namely $S_{ut+i} = P_{i,u}$.
\end{construction}

To illustrate \cnref{wtbadapt}, consider the example in Table\,\ref{tab:wtbadaptex}. Suppose node 1 is erased. Since $S_{1}$ is an invariant subspace of $A_{j}$, for $j \in \{3,5,6,7,8\}$, we can compute \eqref{systematictransmission} for $j \in \{3,5,6,7,8\}$ and $h=1$ if node $j$ sends the symbols $c_{j,0}$ and $c_{j,1}$. Similarly, parity nodes $8+i$ for $i \in \{1,2,3,4\}$ must send the symbols $c_{8+i,0}$ and $c_{8+i,1}$. 

To repair node 1, it remains to describe the symbols that node 2 and node 4 send. Observe that $e_{1}$ is an eigenvector of $A_{2}$, but $e_{0}$ is not. Since $e_{0}+e_{1}+e_{2}+e_{3}$ is an eigenvector of $A_{2}$, we have that $e_{0}A_{2}$, $e_{0}A_{2}^{2}$, $e_{0}A_{2}^{3}$ lie in $\langle e_{0}, e_{1}, e_{2}, e_{3} \rangle$. Consequently, to compute \eqref{systematictransmission} for $h=1$ and $j=2$, it suffices for node $2$ to send the four symbols $c_{2,0}$, $c_{2,1}$, $c_{2,2}$, and $c_{2,3}$. Similarly, to compute \eqref{systematictransmission} for $h=1$ and $j=4$, it suffices for node $4$ to send the four symbols $c_{4,0}$, $c_{4,1}$, $c_{4,2}$ and $c_{4,3}$. 

\tref{wtbadaptbw} shows that the code in \cnref{wtbadapt} has a repair-by-transfer scheme with asymptotically optimal repair bandwidth. 
\begin{theorem}
\label{thm:wtbadaptbw}
The code in \cnref{wtbadapt} has a repair-by-transfer scheme with repair bandwidth at most
\begin{equation}
\label{wtbadaptbwbound}
\left( \frac{(n-1) + 2 \sum\limits_{v=1}^{m-1} (r-s^{v})}{n-k} \right) l,
\end{equation}
which asymptotically meets \eqref{cutset} for fixed $n-k$ as $n \rightarrow \infty$.
\end{theorem}

To verify that the code constructed in \cnref{wtbadapt} can repair systematic nodes with asymptotically optimal repair bandwidth, we check the conditions of \tref{sufficientforAOB}; all proofs are presented in \aref{firsttwo}. Similarly, the full proof of \tref{wtbadaptbw} is presented in \aref{firsttwo}.

\appendix
\section{Proof of \tref{main}}
\label{app:appendixa}

In this section, we prove \tref{main}. We first improve the bound in \clref{case1}. If $S$ is a set and $d \in \mathbb{Z}^{+}$ is a positive integer, recall that $S^{d}$ is the Cartesian product of $S$ taken $d$ times. Define 
\begin{align}
X_{d} &= [0,s-1]^{d} \setminus \{(s-1, \ldots, s-1)\}, \label{Xd} \\
Y_{d} &= [0,s-1]^{d} \setminus \{(0, \ldots, 0)\} \label{Yd}.
\end{align}

\begin{claim}
\label{clm:case1improve}
If $t \in [0,i-1]$ and $i-t \geq m$, then 
\[
\dim_{F}(\{f_{i,j}(\beta^{s^{t}})\}_{j=1}^{l}) \leq \frac{l}{n-k} + \frac{(s^m-1)l}{s^{i-t+m}}.
\]
\end{claim}

\begin{figure*}[t]
\centering
\begin{center}
\begin{tikzpicture}[thick]
  \node[rectangle,draw,label=left:$\mathbf{a}$,label=above:$n+1$](a){*};
  \node[inner sep=0,minimum size=0,right of=a] (b) {$\cdots$}; 
  \node[draw,rectangle,right of=b, label=above:$i+3$] (c) {*};
  \node[rectangle,draw,label=above:$i+2$,right of =c](d){0};
  \node[draw,rectangle,label=above:$i+1$,right of=d] (e) {0};
  \node[draw,rectangle,label=above:$i$,right of=e] (f) {0};
  \node[draw,rectangle,label=above:$i-1$,right of=f] (g) {*};
  \node[draw,rectangle,label=above:$i-2$,right of=g] (h) {*};
  \node[draw,rectangle,label=above:$i-3$,right of=h] (i) {*};
  \node[draw,rectangle,label=above:$i-4$,right of=i] (j) {*};
  \node[draw,rectangle,label=above:$i-5$,right of=j] (k) {*};
  \node[inner sep=0,minimum size=0,right of=k] (l) {$\cdots$}; 
  \node[draw,rectangle,label=above:$0$,right of=l] (m) {*};
  \node[rectangle,draw,label=left:$\mathbf{zs^{t}}$,below of=a](n){0};
  \node[inner sep=0,minimum size=0,below of=b] (o) {$\cdots$}; 
  \node[rectangle,draw,below of=c](p){0};
  \node[rectangle,draw,below of=d](q){0};
  \node[rectangle,draw,below of=e](r){0};
  \node[rectangle,draw,below of=f](s){$z_{2}$};
  \node[rectangle,draw,below of=g](t){$z_{1}$};
  \node[rectangle,draw,below of=h](u){$z_{0}$};
  \node[rectangle,draw,below of=i](v){0};
  \node[rectangle,draw,below of=j](w){0};
  \node[rectangle,draw,below of=k](x){0};
  \node[inner sep=0,minimum size=0,below of=l] (y) {$\cdots$}; 
  \node[rectangle,draw,below of=m](z){0};
  \node[rectangle,draw,label=left:$\mathbf{u}$,below of=n](aa){*};
  \node[inner sep=0,minimum size=0,below of=o] (bb) {$\cdots$}; 
  \node[rectangle,draw,below of=p](cc){*};
  \node[rectangle,draw,below of=q](dd){0};
  \node[rectangle,draw,below of=r](ee){1};
  \node[rectangle,draw,below of=s](ff){0};
  \node[rectangle,draw,below of=t](gg){?};
  \node[rectangle,draw,below of=u](hh){?};
  \node[rectangle,draw,below of=v](ii){*};
  \node[rectangle,draw,below of=w](jj){*};
  \node[rectangle,draw,below of=x](kk){*};
  \node[inner sep=0,minimum size=0,below of=y] (ll) {$\cdots$}; 
  \node[rectangle,draw,below of=z](mm){*};
\end{tikzpicture}
\caption{The $s$-ary expansions of $a$, $zs^{t}$, and $u = a+zs^{t}$ in \clref{case2} when $m=3$, $t=i-2$, and $u \in S_{i,t,1}$}
\label{fig:case2}
\end{center}
\end{figure*}

\begin{figure*}[t]
\centering
\begin{center}
\begin{tikzpicture}[thick]
  \node[rectangle,draw,label=left:$\mathbf{a}$,label=above:$n+2$](a){0};
  \node[rectangle,draw,label=above:$n+1$,right of =a](b){*};
  \node[inner sep=0,minimum size=0,right of=b] (c) {$\cdots$}; 
  \node[rectangle,draw,label=above:$i+6$,right of =c](d){*};
  \node[draw,rectangle,label=above:$i+5$,right of=d] (e) {*};
  \node[draw,rectangle,label=above:$i+4$,right of=e] (f) {*};
  \node[draw,rectangle,label=above:$i+3$,right of=f] (g) {*};
  \node[draw,rectangle,label=above:$i+2$,right of=g] (h) {0};
  \node[draw,rectangle,label=above:$i+1$,right of=h] (i) {0};
  \node[draw,rectangle,label=above:$i$,right of=i] (j) {0};
  \node[draw,rectangle,label=above:$i-1$,right of=j] (k) {*};
  \node[inner sep=0,minimum size=0,right of=k] (l) {$\cdots$}; 
  \node[draw,rectangle,label=above:$0$,right of=l] (m) {*};
  \node[rectangle,draw,label=left:$\mathbf{zs^{t}}$,below of=a](n){0};
  \node[rectangle,draw,below of=b](o){0};
  \node[inner sep=0,minimum size=0,below of=c] (p) {$\cdots$}; 
  \node[rectangle,draw,below of=d](q){0};
  \node[rectangle,draw,below of=e](r){0};
  \node[rectangle,draw,below of=f](s){$z_{2}$};
  \node[rectangle,draw,below of=g](t){$z_{1}$};
  \node[rectangle,draw,below of=h](u){$z_{0}$};
  \node[rectangle,draw,below of=i](v){0};
  \node[rectangle,draw,below of=j](w){0};
  \node[rectangle,draw,below of=k](x){0};
  \node[inner sep=0,minimum size=0,below of=l] (y) {$\cdots$}; 
  \node[rectangle,draw,below of=m](z){0};
  \node[rectangle,draw,label=left:$\mathbf{u}$,below of=n](aa){1};
  \node[rectangle,draw,below of=o](bb){0};
  \node[inner sep=0,minimum size=0,below of=p] (cc) {$\cdots$}; 
  \node[rectangle,draw,below of=q](dd){0};
  \node[rectangle,draw,below of=r](ee){0};
  \node[rectangle,draw,below of=s](ff){?};
  \node[rectangle,draw,below of=t](gg){?};
  \node[rectangle,draw,below of=u](hh){$z_{0}$};
  \node[rectangle,draw,below of=v](ii){0};
  \node[rectangle,draw,below of=w](jj){0};
  \node[rectangle,draw,below of=x](kk){*};
  \node[inner sep=0,minimum size=0,below of=y] (ll) {$\cdots$}; 
  \node[rectangle,draw,below of=z](mm){*};
\end{tikzpicture}
\caption{The $s$-ary expansions of $a$, $zs^{t}$, and $u = a+zs^{t}$ in \clref{case3} when $m=3$, $t=i+2$, and $u \in S_{i,t,1}$}
\label{fig:case3}
\end{center}
\end{figure*}

\noindent \textbf{Proof.} Define the set
\begin{align}
S_{i,t} = \{&u \in [0,l-1] : \, u_{i+m-1} = \cdots = u_{i+1} = 0, \notag \\ 
&u_{i} = 1, u_{i-1} = \cdots = u_{t+m} = 0, \label{carries1improve} \\ 
&(u_{t+m-1}, \ldots, u_{t}) \in X_{m}\}. \label{carriesinitial1} 
\end{align}
We claim that if $t \in [0,i-1]$ and $i-t \geq m$, then 
\begin{align*}
\{f_{i,j}(\beta^{s^{t}})\}_{j=1}^{l} &= \{\beta^{a+zs^{t}} : a \in S_{i}, z \in [0,n-k-1]\} \notag \\ 
&\subseteq \{\beta^{u} : u \in S_{i} \cup S_{i,t}\}.
\end{align*}
As in the proof of \clref{case1}, if $u = a+zs^{t} \notin S_{i}$, then the addition must have generated a carry from coordinate $t+m-1$ to coordinate $i$, which motivates \eqref{carries1improve}. Moreover, let $c=zs^{t}$ and let $y \in [t,t+m-1]$ denote the first coordinate from which there is a carry in $u$. Since $a_{y}$ and $c_{y}$ lie in $[0,s-1]$, this implies that $u_{y} \in [0,s-2]$ and that \eqref{carriesinitial1} holds. Consequently, if $u \notin S_{i}$ then $u \in S_{i,t}$. 
\clref{case1improve} follows because
\begin{equation*}
|S_{i}| = \frac{l}{n-k} \, \, \, \mbox{and} \, \, \, |S_{i,t}| = \frac{(s^m-1)l}{s^{i-t+m}}. \qed
\end{equation*}

In \tref{weakmain}, we proved a weaker version of \tref{main} by using, for all $t \in [i-(m-1),i+(m-1)]$, the trivial bound \eqref{weakcase2and3}. We now improve \eqref{weakcase2and3} in this case.

\begin{claim}
\label{clm:case2}
If $t \in [0,i-1]$ and $i-t < m$, write $t = i-w$ where $w \in [m-1]$. We have 
\[
\dim_{F}(\{f_{i,j}(\beta^{s^{t}})\}_{j=1}^{l}) \leq \frac{l}{n-k} + \frac{(s^{m}-1)l}{s^{m+w}}.
\]
\end{claim}

\noindent \textbf{Proof.} Define the sets
\begin{align}
S_{i,t,1} = \{u \in [0,l-1] : \, &u_{i+m-1} = \cdots = u_{i+m-w+1} = 0, \notag \\ 
&u_{i+m-w} = 1, \notag \\ 
&u_{i+m-w-1} = \cdots = u_{i} = 0, \notag \\
&(u_{i-1}, \ldots, u_{i-w}) \in X_{w}\}, \label{carriesinitial2} 
\end{align}
\begin{align*}
S_{i,t,0} = \{u \in [0,l-1] : \, &u_{i+m-1} = \cdots = u_{i+m-w} = 0, \notag \\ 
&(u_{i+m-w-1}, \ldots, u_{i}) \in Y_{m-w}\}. 
\end{align*}

We claim that if $t \in [0,i-1]$ and $i-t < m$, then 
\begin{align*}
\{f_{i,j}(\beta^{s^{t}})\}_{j=1}^{l} &= \{\beta^{a+zs^{t}} : a \in S_{i}, z \in [0,n-k-1]\} \notag \\ 
&\subseteq \{\beta^{u} : u \in S_{i} \cup S_{i,t,1} \cup S_{i,t,0} \}.
\end{align*}
Please refer to Figure~\ref{fig:case2} and let $u = a+zs^{t}$. By considering the $s$-ary expansions of $a$ and $zs^{t}$, we have $u_{i+m-w} \in \{0,1\}$. In order for $u_{i+m-w} = 1$, we need 
\begin{equation*}
u_{i} = \cdots = u_{i+m-w-1} = 0 
\end{equation*}
and carries from coordinate $i-1$ to coordinate $i+m-w$. As in \clref{case1improve}, we must have that \eqref{carriesinitial2} holds because the position of first carry occurs in $[i-w,i-1]$. Hence, if $u_{i+m-w}=1$ then $u \in S_{i,t,1}$. 

If $u_{i+m-w}=0$ and $u \notin S_{i}$ then $u \in S_{i,t,0}$. \clref{case2} holds because
\begin{equation*}
|S_{i,t,1}| = \frac{(s^{w}-1)l}{s^{m+w}}, \, \, \, \mbox{and} \, \, \, |S_{i,t,0}| = \frac{(s^{m}-s^{w})l}{s^{m+w}}. \qed
\end{equation*}

\begin{claim}
\label{clm:case3}
If $t \in [i+1,n-1]$ and $t-i < m$, write $t = i+w$ where $w \in [m-1]$. We have 
\[
\dim_{F}(\{f_{i,j}(\beta^{s^{t}})\}_{j=1}^{l}) \leq \frac{l}{n-k} + \frac{(s^{w}-1)l}{s^{n+w-i-1}} + \frac{(s^{m-w}-1)l}{s^{m}}.
\]
\end{claim}

\noindent \textbf{Proof.} Define the sets
\begin{align}
S_{i,t,1} = \{u \in \, &[0,s^{m+n}-1] : u_{m+n-1} = 1, \notag \\ 
&u_{m+n-2} = \cdots = u_{m+w+i} = 0, \notag \\
&(u_{i+m+w-1}, \ldots, u_{i+m}) \in X_{w} \label{carriesinitial3} \\
&u_{i+w-1} = \cdots = u_{i} = 0\}, \notag 
\end{align}
\begin{align*}
S_{i,t,0} = \{u \in [0,l-1] : \,&(u_{i+m-1}, \ldots, u_{i+w}) \in Y_{m-w}, 
\\ 
&u_{i+w-1} = \cdots = u_{i} = 0\}. 
\end{align*}
We claim that if $t \in [i+1,n-1]$ and $t-i < m$, then 
\begin{align*}
\{f_{i,j}(\beta^{s^{t}})\}_{j=1}^{l} &= \{\beta^{a+zs^{t}} : a \in S_{i}, z \in [0,n-k-1]\} \notag \\ 
&\subseteq \{\beta^{u} : u \in S_{i} \cup S_{i,t,1} \cup S_{i,t,0} \}.
\end{align*}
Please refer to Figure~\ref{fig:case3} and let $u = a+zs^{t}$. By considering the $s$-ary expansions of $a$ and $zs^{t}$, we have $u_{m+n-1} \in \{0,1\}$. In order for $u_{m+n-1} = 1$, we need 
\begin{equation*}
u_{m+w+i} = \cdots = u_{m+n-2} = 0 
\end{equation*}
and carries from the coordinate $i+w+m-1$ to the coordinate $m+n-1$. As in \clref{case1improve}, we have that \eqref{carriesinitial3} holds because the position of first carry lies in the interval $[i+m, i+w+m-1]$. Hence, if $u_{n+m-1}=1$ then we must have $u \in S_{i,t,1}$. 

If $u_{n+m-1}=0$ and $u \notin S_{i}$ then $u \in S_{i,t,0}$. \clref{case3} holds because
\begin{equation*}
|S_{i,t,1}| = \frac{(s^{w}-1)l}{s^{n+w-i-1}} , \, \, \, \mbox{and} \, \, \, |S_{i,t,0}| = \frac{(s^{m-w}-1)l}{s^{m}}. \qed
\end{equation*}

Finally, we bound \eqref{sumofdimensions} from above by \eqref{bwbound}. Let 
\begin{equation*}
M = \min\{m-1, n-1-i\}
\end{equation*}
denote the minimum of $m-1$ and $n-1-i$. Summing the bounds in \clref{case1improve}, \clref{case2}, \clref{case3}, and  \clref{case4} respectively, an upper bound on \eqref{sumofdimensions} is
\begin{align}
\frac{(n-1)l}{n-k} &+ \sum_{t=1}^{i} \frac{(s^{m}-1)l}{s^{m+t}} + \sum_{w=1}^{M} \frac{(s^{w}-1)l}{s^{n+w-i-1}} \notag \\ 
&+ \sum_{w=1}^{M} \frac{(s^{m-w}-1)l}{s^{m}} + \sum_{t=i+m}^{n-1} \frac{l}{s^{m+n-t-1}}. \label{sumupcases}
\end{align}
If $i > n-1-m$, then the last sum in \eqref{sumupcases} is omitted. 

The first sum in \eqref{sumupcases} comes from combining the bounds in \clref{case1improve} and \clref{case2}, and holds regardless of whether $i < m$ or $i \geq m$. Using well-known formulas for geometric series, we can bound the first sum in \eqref{sumupcases} from above by 
\begin{equation}
\label{case1and2improve}
\sum_{t=1}^{i} \frac{(s^{m}-1)l}{s^{m+t}} \leq \frac{(s^{m-1} + \cdots + 1)l}{s^m}.
\end{equation}

Note that the second sum in \eqref{sumupcases} only appears if $i < n-1$. Hence, we can bound the second sum in \eqref{sumupcases} from above by
\begin{equation}
\label{case3sum}
\sum_{w=1}^{M} \frac{(s^{w}-1)l}{s^{n+w-i-1}} \leq \frac{Ml}{s^{n-i-1}} < \frac{l}{s}. 
\end{equation}
Similarly, we can bound the third sum in \eqref{sumupcases} by
\begin{equation}
\label{case3sumpart2}
\sum_{w=1}^{M} \frac{(s^{m-w}-1)l}{s^{m}} \leq \frac{(s^{m-1} + \cdots + s - (m-1))l}{s^{m}}.
\end{equation}

Finally, using well-known formulas for geometric series, we can bound the last sum in \eqref{sumupcases} by \eqref{case4sum}. Summing the right-hand-sides of \eqref{case1and2improve}, \eqref{case3sum}, \eqref{case3sumpart2}, and \eqref{case4sum} yields that \eqref{sumofdimensions} is bounded above by \eqref{bwbound}. This completes the proof of \tref{main}. \qed


\section{Proof of \tref{bb}}
\label{app:appendixb}

In this section, we prove \tref{bb}. We proved a weaker version of \tref{bb} in \tref{weakbb} by allowing, in the repair of node $i$, for node $j$ to transmit all of its elements if $|j-i| < m$. We now show that node $i$ can still be repaired if node $j$ transmits only $s^{|j-i|}$ scalars in $F$ when $|j-i| < m$. The proof of \tref{bb} uses notation defined in the proof of \tref{weakbb}. In conjunction with the proof of \tref{bb}, the reader may find it useful to consult \eref{exbb}, which illustrates the notation of \tref{bb}.

\vspace{0.25cm}

\noindent \textbf{Proof of \tref{bb}.}
For integers $a \in [0,l-1]$ and $i \in [n]$, we will show that the $r$ coordinates in \eqref{rcoords} are functions of a set of at most $n-1 + 2 \sum_{v=1}^{m-1} (s^{v}-1)$ elements of $F$.

For $a \in [0,l-1]$, $j \in [n] \setminus \{i\}$ such that $|j-i| = w < m$, and $(b_{m-w}, \ldots, b_{1}) \in [0,s-1]^{m-w}$, we must first define a set $T_{a,j,i}(b_{m-w}, \ldots, b_{1})$. If $j>i$, then $T_{a,j,i}(b_{m-w}, \ldots, b_{1})$ is the set of all integers in $[0,l-1]$ whose $s$-ary representation is obtained from $a$ by replacing the $m$ coordinates $a_{i+m-1}, \ldots, a_{i}$ with $b_{m-w}, \ldots, b_{1}, d_{w}, \ldots, d_{1}$ respectively for all tuples $(d_{w}, \ldots, d_{1}) \in [0, s-1]^{w}$. If $j < i$, then $T_{a,j,i}(b_{m-w}, \ldots, b_{1})$ is the set of all integers in $[0,l-1]$ whose $s$-ary representation is obtained from $a$ by replacing the $m$ coordinates $a_{i+m-1}, \ldots, a_{i}$ with $d_{m}, \ldots, d_{m-w+1}, b_{m-w}, \ldots, b_{1}$ respectively for all $(d_{m}, \ldots, d_{m-w+1}) \in [0,s-1]^{w}$; that is
\begin{align}
\label{Tjdef}
&T_{a,j,i}(b_{m-w}, \ldots, b_{1}) \notag \\
&= \begin{cases}
\{a(i; b_{m-w}, \ldots, b_{1}, d_{w}, \ldots, d_{1}) \\
: (d_{w}, \ldots, d_{1}) \in [0, s-1]^{w}\} & \text{if $j>i$} \\
\{a(i; d_{m}, \ldots, d_{m-w+1}, b_{m-w}, \ldots, b_{1}) \\
: (d_{m}, \ldots, d_{m-w+1}) \in [0,s-1]^{w} \} & \text{if $j<i$.} 
\end{cases}
\end{align}
For $(b_{m-w}, \ldots, b_{1}) \in [0,s-1]^{m-w}$, $a \in [0,l-1]$, and $j \in [n] \setminus \{i\}$, we define elements in $F$ by 
\begin{equation}
\label{ujiab}
u^{(a)}_{j,i}(b_{m-w}, \ldots, b_{1}) = \sum_{a' \in T_{a,j,i}(b_{m-w}, \ldots, b_{1})} c_{j,a'}.
\end{equation}

Recall the definition of $u^{(a)}_{j,i}$ from \eqref{ujia}. We claim that the $r$ coordinates in \eqref{rcoords} are functions of the set $E^{(a)}_{i} \subset F$,
\begin{align*}
E^{(a)}_{i} &= \{u^{(a)}_{j,i} : |j-i| \geq m\} \; \sqcup \{u^{(a)}_{j,i}(b_{m-w}, \ldots, b_{1}) : \notag \\
&|j-i| = w < m, (b_{m-w}, \ldots, b_{1}) \in [0,s-1]^{m-w} \}.
\end{align*}
It may be useful to consult \eref{exbb}, which illustrates this claim for $a=0$ and $i=2$ in our running example of $s=2$, $m=2$, $r=4$, $n=10$, and $l=2^{11}$. To repair $r$ coordinates in the failed node $i$, a surviving node $j$ needs to transmit one scalar in $F$ if $|j-i| \geq m$ and $s^{|j-i|}$ scalars in $F$ otherwise. Consequently, the size of $E^{(a)}_{i}$ is at most $n-1 + 2 \sum_{v=1}^{m-1} (s^{v}-1)$, and so the repair bandwidth of the code in \cnref{adapt} is at most \eqref{bandwidthbound}. 

We write \eqref{coordinatewise} for $t \in [0,r-1]$ and sum over $a' \in S_{a,i}$. When $t=0$, we still obtain \eqref{tis0}. If $|i-j| = w < m$, then the value $a'(j+m-1, \ldots, j)$ is the same  over all $a'$ in $T_{a,j,i}(b_{m-w}, \ldots, b_{1})$ for a fixed $(b_{m-w}, \ldots, b_{1})$ in $[0,s-1]^{m-w}$. For $j$ satisfying $|i-j|=w < m$ and a fixed $(b_{m-w}, \ldots, b_{1}) \in [0,s-1]^{m-w}$, define $l_{a,j}(b_{m-w}, \ldots, b_{1}) \in [0,r-1]$ to be the value of $a'(j+m-1, \ldots, j)$ for any $a' \in T_{a,j,i}(b_{m-w}, \ldots, b_{1})$. Recalling the definition of $l_{a,j}$ from the proof of \tref{weakbb}, we obtain when $1 \leq t \leq r-1$ that 
\begin{align}
\label{tisnotzero}
&\sum_{a' \in S_{a,i}} \lambda^{t}_{i, a'_{(i+m-1, \ldots, i)}} c_{i, a'} = - \sum_{j : |j-i| \geq m} \lambda^{t}_{j, l_{a,j}} u^{(a)}_{j,i} \notag \\
&- \sum_{j : |j-i| = w < m} \sum_{\substack{(b_{m-w}, \ldots, b_{1}) \\ \in [0,s-1]^{m-w}}} \lambda^{t}_{j, l_{a,j}(b_{m-w}, \ldots, b_{1})}u^{(a)}_{j,i}(b_{m-w}, \ldots, b_{1}).
\end{align}

For $t=0$, let $B_{0}$ denote the right-hand-side of \eqref{tis0}. Similarly, for $1 \leq t \leq r-1$, let $B_{t}$ denote the right-hand-side of \eqref{tisnotzero}. Since $B_{0}, \ldots, B_{r-1}$ are functions of the elements in $E^{(a)}_{i}$ and since $\lambda_{i,0}, \ldots, \lambda_{i,r-1}$ are distinct, we can solve the system in \eqref{matrixform} for the $r$ coordinates in \eqref{rcoords} given $E^{(a)}_{i}$. \qed

To illustrate \tref{bb}, consider the following example.

\vspace{0.1cm}

\begin{example}
\label{ex:exbb}
Suppose $s=2$, $m=2$, $r=4$, $n=10$, and $l=2^{11}$. Suppose node 2 has failed so $i=2$. Letting $a=0$, we have from \eref{exweakbb} that $S_{0,2} = \{0,2,4,6\}$ and the four coordinates in \eqref{rcoords} are $\{c_{2,0}, c_{2,2}, c_{2,4}, c_{2,6}\}$. 
\end{example}

In this example, equations \eqref{Tjdef} and \eqref{ujiab} are
\begin{align*}
T_{0,1,2}(0) &= \{0,4\}, \; T_{0,1,2}(1) = \{2,6\}, \; \notag \\
T_{0,3,2}(0) &= \{0,2\}, \; T_{0,3,2}(1) = \{4,6\},  
\\
u^{(0)}_{1,2}(0) &= c_{1,0} + c_{1,4}, \; u^{(0)}_{1,2}(1) = c_{1,2} + c_{1,6}, \notag \\ 
u^{(0)}_{3,2}(0) &= c_{3,0} + c_{3,2}, \; u^{(0)}_{3,2}(1) = c_{3,4} + c_{3,6}. 
\end{align*}

Recalling the definition of $u^{(0)}_{j,2}$ from \eqref{ujiaex}, we claim these coordinates are functions of the set
\begin{align}
&E^{(0)}_{2} = \{ u^{(0)}_{j,2} : 4 \leq j \leq 10 \} \; \cup \notag \\
&\{u^{(0)}_{1,2}(0), u^{(0)}_{1,2}(1), u^{(0)}_{3,2}(0), u^{(0)}_{3,2}(1) \}. \label{thesetexample}
\end{align}

To see why the coordinates $c_{2,0}, c_{2,2}, c_{2,4}, c_{2,6}$ are functions of the set $E^{(0)}_{2}$ in \eqref{thesetexample}, sum the equations in \eref{excoordinatewise}. When $t = 0$, we still obtain \eqref{tis0ex}. We have $l_{0,j} = 0$ for $4 \leq j \leq 10$, $l_{0,1}(0) = 0$, $l_{0,1}(1) = 2$, $l_{0,3}(0) = 0$, and $l_{0,3}(1) = 1$, so when
$1 \leq t \leq 3$, we obtain
\begin{align*}
&\lambda^{t}_{2,0}c_{2,0} + \lambda^{t}_{2,1}c_{2,2} + \lambda^{t}_{2,2}c_{2,4} + \lambda^{t}_{2,3}c_{2,6} \notag \\
&= -\lambda^{t}_{1,0}(c_{1,0} + c_{1,4}) - \lambda^{t}_{1,2}(c_{1,2} + c_{1,6}) - \lambda^{t}_{3,0}(c_{3,0} + c_{3,2}) \notag \\
&- \lambda^{t}_{3,1}(c_{3,4} + c_{3,6}) -\sum_{j=4}^{10} \lambda^{t}_{j,0} (c_{j,0} + c_{j,2} + c_{j,4} + c_{j,6}) \notag \\
&= -\lambda^{t}_{1,0}u^{(0)}_{1,2}(0) - \lambda^{t}_{1,2}u^{(0)}_{1,2}(1) - \lambda^{t}_{3,0}u^{(0)}_{3,2}(0) - \lambda^{t}_{3,1}u^{(0)}_{3,2}(1) \notag \\
&-\sum_{j=4}^{10} \lambda^{t}_{j,0} u^{(0)}_{j,2}.
\end{align*}


\section{Verifying \cnref{wtbadapt} and Proof of \tref{wtbadaptbw}}
\label{app:firsttwo}

To show that the code constructed in \cnref{wtbadapt} can repair systematic nodes with asymptotically optimal repair bandwidth, we check the conditions of \tref{sufficientforAOB}. We have specified the left eigenspaces of the encoding matrix $A_{j}$ in \cnref{wtbadapt}. We claim that these eigenspaces each have dimension $l/r$ and that, for $u \in [0,r]$, any $r$ eigenspaces of the form $P_{i,u}$ span $F^{l}$, as required in the first condition of \tref{sufficientforAOB}

\begin{lemma}
\label{lem:dimandbasis}
For $i \in [t]$ and $u \in [0,r]$, we have $\dim(P_{i,u}) = l/r$. Moreover, for any $u' \in [0,r]$, we have 
\begin{equation}
\label{basis}
\dim \left( \bigvee_{\substack{u = 0 \\ u \neq u'}}^{r} P_{i,u} \right) = l.
\end{equation}
\end{lemma}

\noindent \textbf{Proof.} Since $r = s^{m}$, $l = s^{m+t-1}$, and $m$ coordinates of $a$ are fixed, by \dref{Piu}, there are $l/r$ standard basis vectors $e_{a}$ in $P_{i,u}$ for $u \in [0,r-1]$. Hence, for $u \in [0,r-1]$, we have $\dim(P_{i,u}) = l/r$. Observe that $P_{i,r}$ is spanned by sums of standard basis vectors, which we will refer to as \textsl{basis sums}. For any fixed $u' \in [0,r-1]$, observe that in $P_{i,r}$, each $a' \in [0,l-1]$ satisfying $a'_{(i,\ldots, i+m-1)} = u'$ appears exactly once in each basis sum. The remaining $e_{a'}$ (those with $a'_{(i,\ldots, i+m-1)} \neq u'$) in each basis sum belong to a subspace in $\{P_{i,0}, \ldots, P_{i,r-1} \} \setminus \{P_{i,u'}\}$. Therefore, $\dim(P_{i,r}) = l/r$ and \eqref{basis} holds for any $u' \in [0,r]$. \qed



We have demonstrated the first two conditions of \tref{sufficientforAOB} by defining the encoding matrices and the repairing subspaces in \cnref{wtbadapt} and by checking that the left eigenspaces of the encoding matrices have the required properties in \lref{dimandbasis}. We now turn to the task of demonstrating the third condition of \tref{sufficientforAOB}, namely proving that, for each $h \in [k]$, the repairing subspace $S_{h}$ is an invariant subspace of $A_{j}$ for all but a constant number of nodes $j \in [k] \setminus \{h\}$. \lref{iisi'}, \lref{separationbym} and \lref{nooverlap} give sufficient conditions for the repairing subspace $S_{ut+i}$ to be an invariant subspace of the encoding matrix $A_{u't+i'}$.

We will show that, for distinct integers $ut+i$, $u't+i' \in [k]$, where $u,u' \in [0,r-1]$ and $i,i' \in [t]$, the repairing subspace $S_{ut+i}$ is an invariant subspace of $A_{u't+i'}$ when $i=i'$, $|i-i'| \geq m$, or $|i-i'| < m$ and for all $a \in [0,l-1]$ such that $a_{(i,\ldots, i+m-1)} = u$, we have $a_{(i', \ldots, i'+m-1)} \neq u'$. The proofs are similar to the corresponding cases in \cite[Theorem 1]{WTBExplicit}.

\begin{lemma}
\label{lem:iisi'}
If $i=i'$ and $u \neq u'$, then $S_{ut+i}A_{u't+i'} = S_{ut+i}$.
\end{lemma}

\noindent \textbf{Proof.}  By \cnref{wtbadapt}, the left eigenspaces of $A_{u't+i}$ are $\{P_{i,0}, \ldots, P_{i,r} \} \setminus \{P_{i,u'}\}$. Since $u \neq u'$, we have that $P_{i,u}$ is a left eigenspace of $A_{u't+i}$ so
\begin{equation*}
\label{iisi'}
S_{ut+i}A_{u't+i} = P_{i,u}A_{u't+i} = P_{i,u} = S_{ut+i}. \qed
\end{equation*}

Now, we show that $S_{ut+i}$ is an invariant subspace of $A_{u't+i'}$ when $|i-i'| \geq m$.

\begin{lemma}
\label{lem:separationbym}
If $|i-i'| \geq m$, then $S_{ut+i}A_{u't+i'} = S_{ut+i}$.
\end{lemma}

\noindent \textbf{Proof.}
We claim that
\begin{equation}
\label{sumint}
P_{i,u} = \sum_{\substack{u'' = 0 \\ u'' \neq u'}}^{r} \left( P_{i,u} \cap P_{i',u''} \right).
\end{equation}
If $u'' \in [0, r-1] \setminus \{u'\}$ then
\begin{equation}
\label{easyint}
P_{i,u} \cap P_{i',u''} = \mathsf{span} \{ e_{a} : a_{(i,\ldots,i+m-1)} = u, a_{(i',\ldots,i'+m-1)} = u'' \}.
\end{equation}
Since $|i-i'| \geq m$, there are $2m$ coordinates of $a$ fixed in \eqref{easyint}. As $r = s^{m}$ and $l = s^{m+t-1}$, there are $l/r^{2}$ standard basis vectors $e_{a}$ in \eqref{easyint}. Thus, $\dim(P_{i,u} \cap P_{i',u''}) = l/r^{2}$ for $u'' \in [0, r-1] \setminus \{u'\}$.

We claim that
\begin{equation*}
P_{i,u} \cap P_{i',r} = \mathsf{span} \left( \sum_{a' \in M_{a,i'}} e_{a'} : a_{(i, \ldots, i+m-1)} = u \right).
\end{equation*}
Notice that if $a_{(i,\ldots,i+m-1)} = u$, then every $a' \in M_{a,i'}$ satisfies $a'_{(i,\ldots,i+m-1)}=u$ since $|i-i'| \geq m$. Hence, if $a_{(i,\ldots,i+m-1)} = u$, then for all $a' \in M_{a,i'}$, we have $e_{a'} \in P_{i,u}$ so 
\begin{equation*}
\sum_{a' \in M_{a,i'}} e_{a'} \in P_{i,u} 
\end{equation*}
and $\dim(P_{i,u} \cap P_{i',r}) = l/r^{2}$. 

Each $e_{a}$ such that $a_{(i,\ldots,i+m-1)}=u$ and $a_{(i',\ldots,i'+m-1)}=u'$ appears exactly once in each basis sum in $P_{i,u} \cap P_{i',r}$. Consequently, since $|i-i'| \geq m$, 
\begin{align*}
P_{i,u} &= \sum_{\hat{u}=0}^{r-1} \mathsf{span} \left(e_{a} : a_{(i,\ldots,i+m-1)}=u, a_{(i',\ldots,i'+m-1)}=\hat{u} \right) \\
&= \left( \sum_{\substack{\hat{u} = 0 \\ \hat{u} \neq u'}}^{r-1} P_{i,u} \cap P_{i', \hat{u}} \right) \vee \left( P_{i,u} \cap P_{i',r} \right) \\
&= \sum_{\substack{u'' = 0 \\ u'' \neq u'}}^{r} (P_{i,u} \cap P_{i',u''}),
\end{align*}
so \eqref{sumint} is verified.

By definition, $S_{ut+i} = P_{i,u}$ so by \eqref{sumint}
\begin{align}
\label{soinvariant}
S_{ut+i}A_{u't+i'} &= P_{i,u}A_{u't+i'} \notag \\
&= \left( \sum_{\substack{u'' = 0 \\ u'' \neq u'}}^{r} P_{i,u} \cap P_{i,u''} \right)A_{u't+i'} \notag \\
&= \sum_{\substack{u'' = 0 \\ u'' \neq u'}}^{r} (P_{i,u} \cap P_{i,u''})A_{u't+i'} \notag \\
&= \sum_{\substack{u'' = 0 \\ u'' \neq u'}}^{r} (P_{i,u} \cap P_{i',u''}) \notag \\
&= P_{i,u} \notag \\
&= S_{ut+i}. \qed
\end{align}

Finally, we show that $S_{ut+i}$ is an invariant subspace of $A_{u't+i'}$ when $0 < |i-i'| < m$ and for all $a \in [0,l-1]$ such that $a_{(i, \ldots, i+m-1)} = u$, we have $a_{(i', \ldots, i'+m-1)} \neq u'$

\begin{lemma}
\label{lem:nooverlap}
If $0 < |i-i'| < m$ and for all $a \in [0,l-1]$ such that $a_{(i, \ldots, i+m-1)} = u$, we have $a_{(i', \ldots, i'+m-1)} \neq u'$ then 
\[
S_{ut+i}A_{u't+i'} = S_{ut+i}.
\]
\end{lemma}

\noindent \textbf{Proof.} Let 
\begin{equation}
\label{Rii'u}
R_{i,i',u} = \left \{ a_{(i', \ldots, i'+m-1)} : a \in [0,l-1] \, \mbox{and} \, a_{(i, \ldots, i+m-1)} = u \right \}
\end{equation}
be the set of all possible values of $a_{(i', \ldots, i'+m-1)} \in [0,r-1]$ given that $a_{(i, \ldots, i+m-1)} = u$. By assumption, $u' \notin R_{i,i',u}$ so
\begin{align}
P_{i,u} &= \mathsf{span} \left( e_{a} : a_{(i, \ldots i+m-1)} = u \right) \notag \\
&= \sum_{x \in R_{i,i',u}} \mathsf{span} \left( e_{a} : a_{(i, \ldots i+m-1)} = u, a_{(i', \ldots i'+m-1)} = x \right) \notag \\
&= \sum_{x \in R_{i,i',u}} (P_{i,u} \cap P_{i',x}) \notag \\
&= \sum_{\substack{u'' = 0 \\ u'' \neq u'}}^{r} (P_{i,u} \cap P_{i',u''}). \label{onlyRii'u}
\end{align}
Therefore, \eqref{sumint} is verified and the proof of the lemma follows from the argument of \eqref{soinvariant}. \qed

We claim that we have now demonstrated the third condition of \tref{sufficientforAOB}, namely that, for each $h \in [k]$, the repairing subspace $S_{h}$ is an invariant subspace of $A_{j}$ for all but a constant number of nodes $j \in [k] \setminus \{h\}$. By \lref{iisi'}, \lref{separationbym}, \lref{nooverlap}, and \eqref{Rii'u}, if $h = ut+i$ for some $u \in [0,r-1]$ and $i \in [t]$, then only nodes $u't+i'$ where $|i-i'| < m$ and $u' \in R_{i,i',u}$ could have $S_{h}$ not be an invariant subspace of $A_{u't+i'}$. For each $i' \in [t]$ with $|i-i'| < m$, there are $|R_{i,i',u}| = s^{|i-i'|}$ nodes $u't+i'$ for which $S_{h}$ may not be an invariant subspace of $A_{u't+i'}$ by \eqref{Rii'u}. Consequently, $S_{h}$ is an invariant subspace of $A_{j}$ for all but at most
\begin{equation}
\label{boundednodes}
2 \sum_{v=1}^{m-1} s^v 
\end{equation}
nodes $j \in [k] \setminus \{h\}$.

We claim that the last condition of \tref{sufficientforAOB} holds. The proof is similar to the proof of the case $i=i'$ and $u=u'$ in \cite[Theorem 1]{WTBExplicit}.

\begin{lemma}
\label{lem:rankconditionrestateholds}
For each $h \in [k]$, \eqref{rankconditionrestate} holds. 
\end{lemma}

\noindent \textbf{Proof.} We only prove the case $u = 0$ as the remaining cases are proved similarly. Denote $A = A_{ut+i}$ and $S_{ut+i} = S$. 

For an integer $a = (a_{1}, \ldots, a_{m+t-1}) \in [0,l-1]$ and integers $i \in [t]$ and $u \in [0,r-1]$, define the integer $a_{i}(u)$ to be the unique element $a' \in M_{a,i}$ with $a'_{(i, \ldots, i+m-1)} = u$. Notice that 
\[
S = \mathsf{span}(P_{i,0}) = \mathsf{span} \{e_{a_{i}(0)} : a \in [0, l-1] \} 
\]
and
\[
P_{i,u} = \mathsf{span} \{e_{a_{i}(u)} : a \in [0, l-1] \} \; \mbox{for $u \in [0,r-1]$}.
\]
Let eigenvalue $\lambda_{0}$ correspond to the eigenspace $P_{i,r}$ and let eigenvalue $\lambda_{\hat{u}}$ correspond to the eigenspace $P_{i, \hat{u}}$ for $\hat{u} \in [1,r-1]$. We then have for $s \in [0,r-1]$ that
\begin{align}
e_{a_{i}(0)}A^{s} &= \left( \left( \sum_{u=0}^{r-1} e_{a_{i}(u)} \right) - e_{a_{i}(1)} - \cdots - e_{a_{i}(r-1)} \right)A^{s} \notag \\
&= \lambda_{0}^{s} \left( \sum_{u=0}^{r-1} e_{a_{i}(u)} \right) - \lambda_{1}^{s} e_{a_{i}(1)} - \cdots - \lambda_{r-1}^{s} e_{a_{i}(r-1)} \notag \\
&= \lambda_{0}^{s} e_{a_{i}(0)} + \sum_{u=1}^{r-1} (\lambda_{0}^{s} - \lambda_{u}^{s}) e_{a_{i}(u)} \label{overand} 
\end{align}
Writing the equations in \eqref{overand} for all $s \in [0,r-1]$ in matrix form yields
\[
\begin{bmatrix}
e_{a_{i}(0)} \\
e_{a_{i}(0)} A \\
\vdots \\
e_{a_{i}(0)} A^{r-1} 
\end{bmatrix}
= M \begin{bmatrix}
e_{a_{i}(0)} \\
e_{a_{i}(1)} \\
\vdots \\
e_{a_{i}(r-1)}  
\end{bmatrix}
\] 
with 
\[
M = \begin{bmatrix}
     1 & 0 & \cdots & 0 \\
     \lambda_{0} & \lambda_{0} - \lambda_{1} & \cdots & \lambda_{0} - \lambda_{r-1} \\
     \lambda_{0}^{2} & \lambda_{0}^{2} - \lambda_{1}^{2} & \cdots & \lambda_{0}^{2} - \lambda_{r-1}^{2} \\
     \vdots & \vdots & \vdots & \vdots \\
     \lambda_{0}^{r-1} & \lambda_{0}^{r-1} - \lambda_{1}^{r-1} & \cdots & \lambda_{0}^{r-1} - \lambda_{r-1}^{r-1}
    \end{bmatrix}.
\]
After a sequence of elementary column operations, $M$ becomes the following Vandermonde matrix
\[
M' = \begin{bmatrix}
     1 & 1 & \cdots & 1 \\
     \lambda_{0} & \lambda_{1} & \cdots & \lambda_{r-1} \\
     \lambda_{0}^{2} & \lambda_{1}^{2} & \cdots & \lambda_{r-1}^{2} \\
     \vdots & \vdots & \vdots & \vdots \\
     \lambda_{0}^{r-1} & \lambda_{1}^{r-1} & \cdots & \lambda_{r-1}^{r-1}
    \end{bmatrix}.
\]
Since $\lambda_{0}, \lambda_{1}, \ldots, \lambda_{r-1}$ are distinct, we know $M'$ and hence $M$ is nonsingular. Therefore,
\[
\mathsf{span} \{ e_{a_{i}(0)}, e_{a_{i}(0)}A, \ldots, e_{a_{i}(0)} A^{r-1} \} = \mathsf{span} \{ e_{a_{i}(0)}, e_{a_{i}(1)}, \ldots, e_{a_{i}(r-1)} \}.  
\]
Since $S$ contains $e_{a_{i}(0)}$ for $a \in [0, l-1]$, we have
\[
S + SA + \cdots + SA^{r-1} = F^{l},
\]
so \eqref{rankconditionrestate} holds. \qed

\subsection{Proof of \tref{wtbadaptbw} and MDS Property}
\label{sec:proofwtbadaptbw}

From \eqref{boundednodes}, we can obtain a weak upper bound on the repair bandwidth of the code in \cnref{wtbadapt}. Let $h = ut+i$ for some $u \in [0,r-1]$ and $i \in [t]$. We can have any node $u't+i'$ not satisfying the hypotheses of \lref{iisi'}, \lref{separationbym}, or \lref{nooverlap} send all of its symbols in the repair of failed node $h$. The number of nodes not satisfying the hypotheses of \lref{iisi'}, \lref{separationbym}, or \lref{nooverlap} is bounded above by \eqref{boundednodes}. The repair bandwidth of the code in \cnref{wtbadapt} is thus at most
\begin{equation}
\label{wtbadaptbwboundweak}
\left( \frac{\left( n-1 - 2 \sum_{v=1}^{m-1} s^v \right) + 2 \left(2 \sum_{v=1}^{m-1} s^v \right)(n-k)}{n-k} \right)l,  
\end{equation}
which asymptotically meets \eqref{cutset} for fixed $n-k$ as $n \rightarrow \infty$.

For $u \in [0,r-1]$, we have $S_{ut+i} = P_{i,u}$, so by \dref{Piu} and the remarks after \tref{sufficientforAOB}, the code in \cnref{wtbadapt} has a repair-by-transfer scheme with asymptotically optimal bandwidth. 

To prove \tref{wtbadaptbw}, we show that if $h = ut+i$ for integers $u \in [0,r-1]$ and $i \in [t]$, then in the repair of failed node $h$, it suffices for any node $u't+i' \in [k] \setminus \{h\}$ that does not satisfy the hypotheses of \lref{iisi'}, \lref{separationbym}, or \lref{nooverlap} to send $l/r + s^{t-|i-i'|-1}(r-2)$ symbols.

\begin{lemma}
\label{lem:fewersymbols}
If $0 < |i-i'| < m$ and there exists $a \in [0,l-1]$ such that $a_{(i, \ldots, i+m-1)} = u$ and $a_{(i', \ldots, i'+m-1)} = u'$, then there exists a set $S_{i,u,i',u'} \subset [0,l-1]$ of size
\begin{equation}
\label{sizeSiui'u'}
\frac{l}{r} + s^{t-|i-i'|-1}(r-s^{|i-i'|})
\end{equation}
such that, to compute \eqref{systematictransmission} for $h = ut+i$ and $j = u't+i'$, it suffices for node $u't+i'$ to send the symbols $\{c_{u't+i', a} : a \in S_{i,u,i',u'}\}$.
\end{lemma}

\noindent \textbf{Proof.}
By \eqref{Rii'u}, we have $u' \in R_{i,i',u}$. Let 
\begin{equation*}
\label{Piuu'}
P_{i,u,u'} = \sum_{x \in R_{i,i',u} \setminus \{u'\}} (P_{i,u} \cap P_{i',x}).
\end{equation*}
For all $x \in R_{i,i',u} \setminus \{u'\}$, the subspace $P_{i',x}$ is a left eigenspace of $A_{u'm+i'}$, so the subspace $P_{i,u,u'}$ is an invariant subspace of $A_{u'm+i'}$. The vectors 
\begin{equation*}
\label{Viui'u'}
V_{i,u,i',u'} = \{e_{a} : a_{(i, \ldots, i+m-1)} = u, a_{(i', \ldots, i'+m-1)} = u'\}
\end{equation*}
lie in $P_{i,u} \cap P_{i',u'}$ and are not eigenvectors of $A_{u'm+i'}$. We have $P_{i,u} = P_{i,i',u} \oplus \mathsf{span}(V_{i,u,i',u'})$ by \eqref{onlyRii'u} and
\begin{equation}
\label{dimViui'u'}
\dim \mathsf{span}(V_{i,u,i',u'}) = s^{(m+t-1)-(m + |i-i'|)} = s^{t-|i-i'|-1}.
\end{equation}

To compute \eqref{systematictransmission} when $h=ut+i$ and $j = u't+i'$, we must compute $e_{a}A_{u't+i'}^{p}$ for $e_{a} \in V_{i,u,i',u'}$ and $p \in [1,r-1]$. Since $e_{a} \in V_{i,u,i',u'}$, we have $\sum_{a' \in M_{a,i'}} e_{a'}$ is an eigenvector of $A_{u't+i'}$. Hence, for $p \in [1,r-1]$, we have $e_{a}A_{u't+i'}^{p} \in \langle e_{a'} : a' \in M_{a,i'} \rangle$. Observe that the set $\{a : e_{a} \in P_{i,u} \}$ is contained in the set $\{a': e_{a} \in V_{i,u,i',u'}, a' \in M_{a,i'} \}$. Hence, to compute \eqref{systematictransmission} when $h=ut+i$ and $j = u't+i'$, it suffices for node $u't+i'$ to send the symbols $\{c_{u't+i', a} : a \in S_{i,u,i',u'}\}$, where
\begin{equation*}
\label{thesymbols}
S_{i,u,i',u} = \{a': e_{a} \in V_{i,u,i',u'}, a' \in M_{a,i'} \}. 
\end{equation*}

To see why the size of $S_{i,u,i',u'}$ is given by \eqref{sizeSiui'u'}, note that if $e_{a}$ and $e_{\hat{a}}$ are distinct elements of $V_{i,u,i',u'}$ then $M_{a,i'}$ is disjoint from $M_{\hat{a},i'}$. By \eqref{dimViui'u'}, the size of $S_{i,u,i',u'}$ is $s^{t-|i-i'|-1}r$, which equals \eqref{sizeSiui'u'}. \qed

We now derive the bound \eqref{wtbadaptbwbound} in \tref{wtbadaptbw}. Recall that, if $h = ut+i$ for some $u \in [0,r-1]$ and $i \in [t]$, then by \lref{iisi'}, \lref{separationbym}, \lref{nooverlap}, and \eqref{Rii'u}, only nodes $u't+i'$ where $|i-i'| < m$ and $u' \in R_{i,i',u}$ do not have $S_{h}$ as an invariant subspace of $A_{u't+i'}$. By \lref{fewersymbols}, each such node must transmit $l/r + s^{t-|i-i'|-1}(r-s^{|i-i'|})$ symbols. Hence, by \eqref{Rii'u} and \eqref{boundednodes}, the repair bandwidth bound is at most
\begin{equation*}
(n-1)\frac{l}{r} + 2 \sum_{v=1}^{m-1} s^{v} s^{t-v-1}(r-s^{v}),
\end{equation*}
which equals \eqref{wtbadaptbwbound} since $s^{t-1} = l/r$.

Finally, we address the MDS property of the code in \cnref{wtbadapt}. By arbitrarily assigning $r$ distinct nonzero eigenvalues to each encoding matrix $A_{j}$, the code in \cnref{wtbadapt} can be made MDS over a large enough field; the proof is the same as the proof of \cite[Theorem 4]{WTBExplicit}.

\vspace{2.00ex}
\section*{Acknowledgement}
We thank Mary Wootters and Hoang Dau for their advice.

\newpage

\bibliographystyle{acm}

\end{document}